\documentclass[11pt]{article}
\usepackage[pdftex]{graphicx}

\usepackage{cite}
\usepackage{amsfonts}
\usepackage{amsmath}
\usepackage{latexsym}
\usepackage{fancybox}
\usepackage{graphicx}

\numberwithin{equation}{section}
\allowdisplaybreaks

\def\spa#1{\phantom{\fbox{\rule[-#1cm]{0cm}{0cm}}}}

\newcommand{\beq}{\begin{equation}}

\newcommand{\eeq}{\end{equation}}
\newcommand{\bea}{\begin{eqnarray}}
\newcommand{\eea}{\end{eqnarray}}

\def\[#1\]{\begin{align}#1\end{align}}

\begin{document}

\hfuzz=100pt
\title{{\Large \bf{
Branched polymers with loops coupled to\\ 
the critical Ising model
}}}
\author{Jan Ambj\o rn$^{a,b}$\footnote{ambjorn@nbi.dk}, 
Yukimura Izawa$^{c,d}$\footnote{izawa-yukimura@hiroshima-u.ac.jp} and Yuki Sato$^{e,f}$\footnote{yukisato@u-fukui.ac.jp} 
  \spa{0.5} \\
\\
$^a${\small{\it The Niels Bohr Institute, Copenhagen University}}
\\ {\small{\it Blegdamsvej 17, DK-2100 Copenhagen, Denmark}}\\    
\\
$^b${\small{\it IMPP, Radboud University}}
\\ {\small{\it Heyendaalseweg 135, 6525 AJ, Nijmegen, The Netherlands}}\\    
\\
$^c${\small{\it Physics Program, Graduate School of Advanced Science and Engineering, Hiroshima University}}
\\ {\small{\it Higashi-Hiroshima 739-8526, Japan}}\\    
\\
$^d${\small{\it International Institute for Sustainability with Knotted Chiral Meta Matter}}
\\ {\small{\it Kagamiyama, Higashihiroshima 739-8511, Hiroshima, Japan}}\\    
\\
$^e${\small{\it Department of Mechanical and System Engineering, University of Fukui}}
\\ {\small{\it 3-9-1 Bunkyo, Fukui-shi, Fukui 910-8507, Japan}}\\    
\\
$^f${\small{\it Department of Physics, Nagoya University}}
\\ {\small{\it Chikusaku, Nagoya 464-8602, Japan}}\\  
\spa{0.3} 
}
\date{}

\maketitle
\centerline{}

\begin{abstract} 
We study the continuum limit of branched polymers (BPs) with loops coupled to Ising spins at the zero-temperature critical point. 
It is known that the continuum partition function can be represented by a Hermitian two-matrix model, and we propose a string field theory whose Dyson{\textendash}Schwinger equation coincides with the loop equation of this continuum matrix model. 

By setting the matrix size to one, we analyze a convergent non-perturbative partition function expressed as a two-dimensional integral, 
and show that it satisfies a third-order linear differential equation. 
In contrast, in the absence of coupling to the critical Ising model, the continuum partition function of pure BPs with loops is known to satisfy the Airy equation.
From the viewpoint of two-dimensional quantum gravity, we introduce a non-perturbative loop amplitude that serves as a solution to the Wheeler{\textendash}DeWitt equation incorporating contributions from all genera. 
Furthermore, we demonstrate that the same Wheeler{\textendash}DeWitt equation can also be derived through the stochastic quantization. 

\end{abstract}

\renewcommand{\thefootnote}{\arabic{footnote}}
\setcounter{footnote}{0}

\newpage

\section{Introduction}
\label{sec:introduction}

Two-dimensional models of quantum geometries provide a useful framework for studying non-perturbative aspects of quantum gravity (see, e.g. Refs.~\cite{Ambjorn:1997di, Ambjorn:2022mxb, Ambjorn:2022btk}).  
In most cases, they can be analyzed analytically through the use of regularization known as dynamical triangulations \cite{Ambjorn:1985az,Ambjorn:1985dn,David:1984tx,Billoire:1985ur,Kazakov:1985ea,Boulatov:1986jd}, or matrix models. 
Branched polymers (BPs) are a simple example of quantum geometries, which are randomly branching graphs without loops. 

Since BPs are essentially random tree graphs, they do not seem to correspond to any extended quantum geometries. 
However, there exist maps between BPs and quantum geometries based on two-dimensional causal dynamical triangulations ($2$D CDT) (see, e.g. Ref.~\cite{Durhuus:2009sm}). 
CDT was first introduced as a model of two-dimensional Lorentzian quantum geometries \cite{Ambjorn:1998xu}, and later extended to models in higher dimensions (see Ref.~\cite{Ambjorn:2024pyv} for recent review). 
Quantum geometries generated by $2$D CDT are extended rather than polymer-like, in which no topology change is allowed to occur. 
In this sense, BPs are related to some extended quantum geometries, 
and the continuum limit of $2$D CDT is known to be the $2$D projectable Ho{\v{r}}ava{\textendash}Litshitz quantum gravity \cite{Ambjorn:2013joa}. 

BPs decorated with loop structures have been extensively studied, and its continuum limit is described by the Airy equation \cite{Nishigaki:1990sk,  Anderson:1991ku, Jurkiewicz:1996yd}. 
The Airy equation is a linear differential equation, and its solutions -- Airy functions -- represent a non-perturbative partition function in the continuum limit. 
Even in this case, a map between BPs with loops and quantum geometries in which the topology change is allowed to occur has been constructed \cite{Ambjorn:2013csx}. 
From the viewpoint of extended quantum geometries, its continuum theory is known as the generalized CDT \cite{Ambjorn:2007jm}, 
and its non-perturbative partition function is essentially given by the Airy functions \cite{Ambjorn:2009fm}. 
The generalized CDT can be formulated in several ways, e.g. a continuous one-matrix model \cite{Ambjorn:2008jf, Ambjorn:2008gk}, a string field theory \cite{Ambjorn:2008ta}, and a stochastic quantization \cite{Ambjorn:2009wi}. 
In this sense, the generalized CDT can be interpreted as the continuum limit of pure BPs with loops
\footnote{An effective gravitational theory corresponding to the generalized CDT can be constructed by introducing a bi-local wormhole interaction to the $2$D projectable Ho{\v{r}}ava{\textendash}Litshitz quantum gravity \cite{Ambjorn:2021wou}.}.        

A two-dimensional model of quantum geometries coupled to the Ising model has been proposed in terms of a Hermitian two-matrix model \cite{Kazakov:1986hu,Boulatov:1986sb}, 
where the Ising model becomes critical at a finite critical temperature. At the finite critical temperature, the continuum theory turns to the Liouville gravity coupled to fermions. 
By slightly deforming the two-matrix model, one can take an unconventional continuum limit in which the branching structure of the graphs becomes important. 
This corresponds to the continuum limit of BPs with loops coupled to the critical Ising model \cite{Sato:2017ccb, Ambjorn:2020kpr}.   
The resulting continuum theory is described by the continuum two-matrix model \cite{Sato:2017ccb}. 
It is known that the Ising model on BPs cannot be critical at any finite temperature \cite{Ambjorn:1992rp}, 
but the criticality discussed in Refs.~\cite{Sato:2017ccb, Ambjorn:2020kpr} occurs at the zero temperature, i.e. quantum criticality. 
In this article, we wish to investigate this continuum theory from various perspectives.  

This article is organized as follows. 
In Section \ref{sec:twomm}, we review how to take the continuum limit of BPs with loops coupled to the critical Ising model in a Hermitian two-matrix model. 
The continuum theory can be expressed again by a certain two-matrix model with renormalized coupling constants. We then derive the corresponding loop equation. 

In Section \ref{sec:sft}, we propose a string field theory for the continuum limit of BPs with loops coupled to the critical Ising model, 
and show the Dyson{\textendash}Schwinger equation precisely reproduces the loop equation derived in the continuum two-matrix model. 

In Section \ref{sec:nonperturbative}, we discuss a non-perturbative partition function by setting $N=1$ in the continuum two-matrix model where $N$ is the size of matrices. 
As it turns out, the non-perturbative partition function satisfies a third-order linear differential equation. 
We also calculate the free energy, and show the loop function that contains contributions from all genera satisfies an integro-differential equation, which is interpreted as the Wheeler{\textendash}DeWitt equation.  

In Section \ref{sec:stochastic}, by identifying the time in the string field theory as the fictitious time in the stochastic quantization, 
we reformulate the continuum theory for BPs with loops coupled to the critical Ising model. Through the Fokker{\textendash}Planck equation, we derive the Wheeler{\textendash}DeWitt equation that is consistent with the one obtained in Section \ref{sec:nonperturbative}.
 
Section \ref{eq:discussions} is devoted to summary and discussions.

\section{Two-matrix model}
\label{sec:twomm}
We consider the following Hermitian two-matrix model \cite{Fuji:2011ce, Sato:2017ccb, Ambjorn:2020kpr}
\[
Z_N(g, c,\theta) 
= \int D\phi_+ D\phi_-\ e^{-\frac{N}{\theta} \text{tr} U(\phi_+, \phi_-)}\ , 
\label{eq:mm}
\]
where $\phi_{\pm}$ are $N\times N$ Hermitian matrices, $D\phi_{\pm}$ the Haar measures on Hermitian matrices, 
and the potential is given by 
\[
U(\phi_+, \phi_-) 
= \frac{1}{2} \left( \phi^2_+ + \phi^2_- \right) 
- c \phi_+ \phi_- 
- g (\phi_+ + \phi_-) 
- \frac{g}{3} \left( \phi^3_+ + \phi^3_- \right)\ . 
\label{eq:potentialU}
\]
Here $g$, $c$, and $\theta$ are non-negative parameters. 
The perturbative expansion of the integral (\ref{eq:mm}) for small $g$ formally defines a model of random graphs consisting of vertices of degree one and three, 
where an Ising spin is assigned to each vertex.

\begin{figure}[h]
\centering
\includegraphics[width=3.5in]{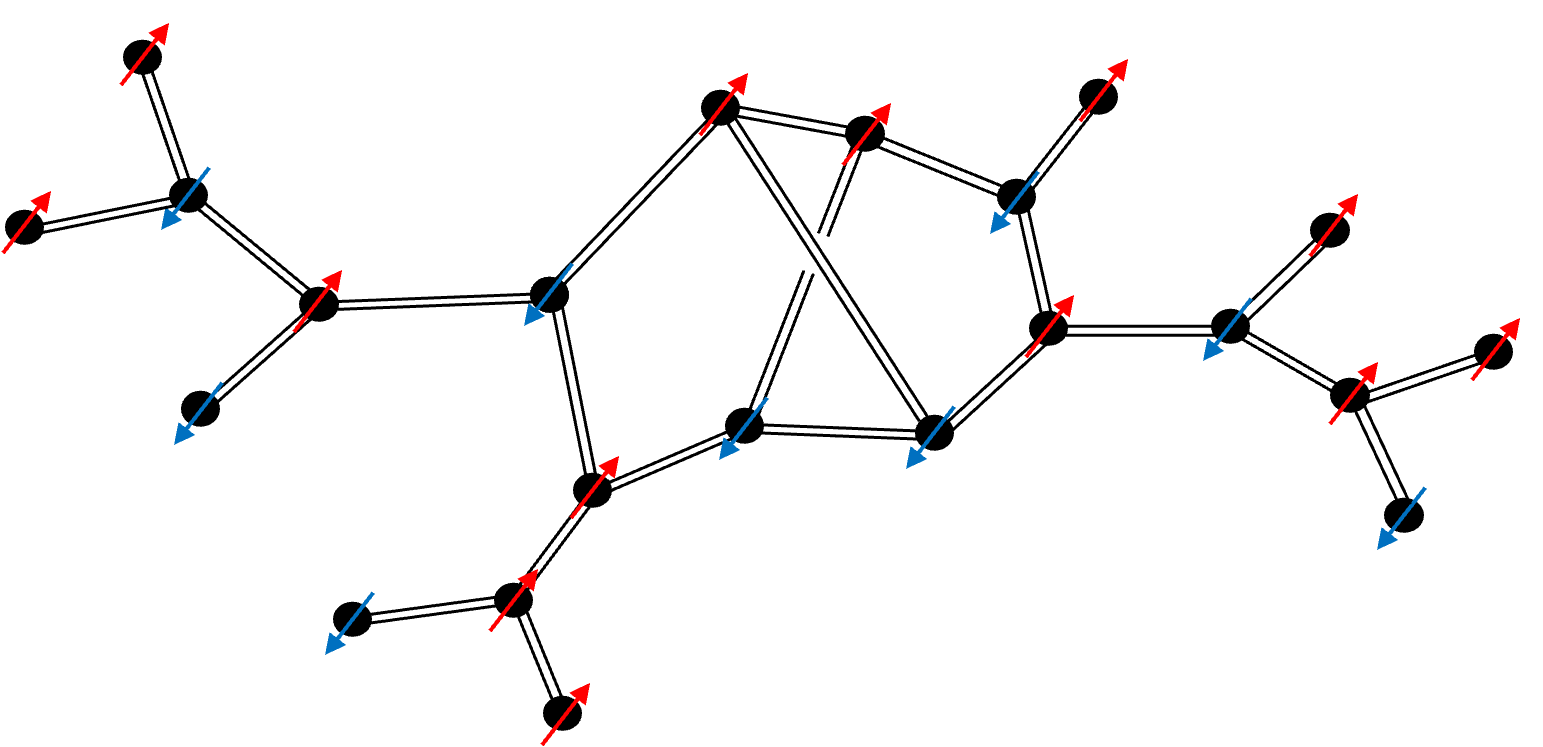}
\caption{A ribbon graph consisting of vertices of degree one and three where the red and blue arrows corresponding to the up and down spins are placed at vertices.}
\label{fig:ribbongraph}
\end{figure}

To demonstrate this, we set $c = e^{-2\beta}$ where $\beta$ is the inverse temperature, and introduce ``propagators''
\[
\left\langle 
\left( \phi_{a} \right)_{ij} \left( \phi_b \right)_{kl}
\right\rangle_0 
=
\frac{\theta}{N} 
\Delta_{ab}  \delta_{il}\delta_{jk}\ , 
\label{eq:propagators}
\]
where the expectation value $\langle \cdot \rangle_0$ is defined in terms of $Z_N(0, c,\theta)$, 
the indices $i,j$ and $a,b$, respectively, ran from $1$ to $N$ and $+$ to $-$, 
and 
\[
\Delta_{ab} 
= 
\frac{\sqrt{c}}{1-c^2}
\begin{pmatrix}
c^{-1/2} & c^{1/2} \\
c^{1/2} & c^{-1/2}
\end{pmatrix} 
= \frac{e^{\beta}}{2 \sinh (2\beta)} 
\begin{pmatrix}
e^{\beta \sigma_+ \sigma_+} & e^{\beta \sigma_+ \sigma_-} \\
e^{\beta \sigma_- \sigma_+} & e^{\beta \sigma_- \sigma_-} 
\end{pmatrix}\ , \ \ \ 
\text{with}\ \ \ 
\sigma_{\pm} = \pm1\ . 
\label{eq:Deltamatrix}
\]
The linear and cubic terms in the potential (\ref{eq:potentialU}) can be associated with the vertices of degree one and three, respectively. 
If we evaluate the perturbative expansion of Eq.~(\ref{eq:mm}) order by order, 
using the propagators (\ref{eq:propagators}),  
we can perform the sum over ribbon graphs consisting of vertices of degree one and three. 
Through the Wick contraction in terms of the propagators (\ref{eq:propagators}), Ising spins are placed at vertices, and they have the nearest neighbor interactions (see Figure \ref{fig:ribbongraph}). 
Note that due to the vertices of degree one, the ribbon graphs generally have the local tree structures. 
As explained later, the parameter $\theta$ is introduced to control such tree structures\footnote{
The parameter $\theta$, together with the linear term in the potential, was first introduced in the context of one-matrix model \cite{Ambjorn:2008gk}, 
aiming for taking the continuum limit of generalized causal dynamical triangulations. 
}. 

\begin{figure}[h]
\centering
\includegraphics[width=3.0in]{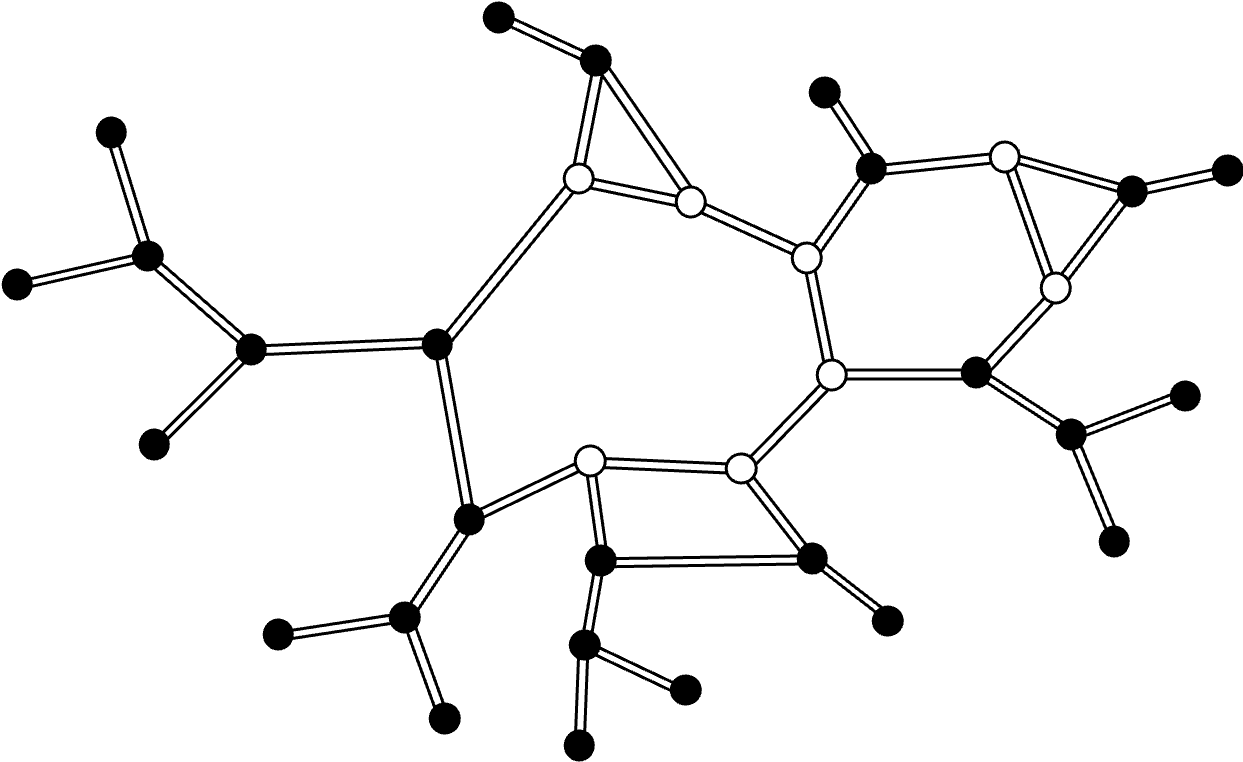}
\caption{Skeleton vertices, shown in white, and black vertices belonging to the local tree structure.}
\label{fig:skeleton}
\end{figure}

Hereafter we consider the connected ribbon graphs. 
For each ribbon graph, the numbers of vertices $V$, edges $E$ and faces $F$ satisfy the topological relation
\[
V-E+F=2-2h\ , 
\label{eq:euler}
\]
where $h$ denotes the genus, i.e. the number of holes. 
Denoting the numbers of vertices of degree one and three by $V_1$ and $V_3$, respectively, we have
\[
2E = 3V_3 + V_1, \ \ \ V = V_3 + V_1\ , \ \ \ V_s = V_3 - V_1\ .
\label{eq:rellationEV}
\]
Here $V_s \ge -2$, and when $V_s > 0$, it denotes the number of skeleton vertices that are the vertices of degree three which do not belong to the local tree structures: 
Let us consider a vertex of degree three and label the three edges emanating from it as $1$, $2$, and $3$ in a clockwise order. 
If, by moving along the edges to other vertices, one can find a path that returns to the original vertex regardless of whether one starts from edge $1$, $2$, or $3$, 
the chosen vertex is referred to as the skeleton vertex\footnote{The back trucking is not allowed.} (See Figure \ref{fig:skeleton}).

Having the discussion above in mind, the sum over connected graphs can be formally expressed as
\[
\frac{1}{N^2} \log \left( 
\frac{Z_N(g,c,\theta)}{Z_N(0,c,\theta)}
\right)
= \sum^{\infty}_{h=0} N^{-2h}
\sum^{\infty}_{V=2} 
\left( \frac{g\sqrt{c}}{1-c^2} \right)^V
\sum^{\infty}_{V_s = -2} 
\left( \frac{\theta \sqrt{c}}{1-c^2} \right)^{\frac{V_s}{2}}
Z_{h,V,V_s} (c)\ , 
\label{eq:connectedsum}
\]
where $Z_{h,V,V_s} (c)$ is the partition function of the Ising model on random graphs for fixed $h$, $V$ and $V_s$. 
As one can see from Eq.~(\ref{eq:connectedsum}), it is $\theta$ that controls the number of skeleton vertices.

In the large-$N$ limit, for fixed $c$ and $\theta$, the sum (\ref{eq:connectedsum}) has the radius of convergence $g_c(\theta, c)$.
Tuning $g$ to $g_c (\theta,c )$ from below, the contribution from infinitely many vertices becomes important in the sum (\ref{eq:connectedsum}), 
and one can take the continuum limit. 
In addition, the Ising model becomes critical at $c=c_c(\theta)$. The critical temperature here is given by $T_c (\theta) = - 2/\log[c_c(\theta)]$.

When $c\ne c_c(\theta)$ with $\theta > 0$, the resulting continuum theory is the pure Liouville gravity. 
At $c=c_c(\theta)$ with $\theta > 0$, the fluctuations of spin variables diverge, and the interaction between geometries and spins becomes strong enough to change the universality class, 
and the continuum theory turns to the Liouville gravity coupled to fermions \cite{Kazakov:1986hu,Boulatov:1986sb}. 
This is true for all $\theta > 0$, and in fact, the average numbers of vertices $\langle V \rangle$ and skeleton vertices $\langle V_s \rangle$ essentially diverge \cite{Ambjorn:2020kpr}.

On the other hand, tuning $\theta \to 0$, along with the critical curve given by $c=c_c (\theta)$,  
one can reach the zero temperature critical point, i.e. the quantum critical point. 
As shown in Ref.~\cite{Ambjorn:2020kpr}, the continuum physics depends on how to approach the zero temperature.  
When we introduce the parametrization,
\[
g=g_c(\theta) e^{- \frac{1}{2} \lambda_0 \varepsilon^2  }\ , \ \ \ 
c=c_c (\theta)\ , \ \ \ \theta = g_s \varepsilon^{a}\ , \ \ \ \text{with}\ \ \ 0< a \le 3\ , 
\label{eq:alphascaling}
\]
where $\varepsilon$ is a lattice spacing, $\lambda_0$ a renormalized cosmological constant, $g_s$ a dimensional constant, 
and $g_c (\theta) = g_c (\theta, c_c(\theta))$ given by 
\[
g_c(\theta) 
=  
\left(
- \frac{9}{4 \times 10^{2/3}} \theta^{2/3} 
+ \frac{ 3^{1/3} \theta^{1/3} (243\theta - 80) + h^2 }{4 \times 30^{2/3} h}
\right)^{3/2}\ , \ \ \ 
c_c (\theta) = 
\left(
\frac{\theta}{10} g^{2}_{c} (\theta)
\right)^{1/3}\ ; 
\label{eq:gc} 
\]
\[
h = 
\left(
81 ( 40 - 81 \theta ) \theta 
+ 80 \left( 
90 + \sqrt{ 8100 + 3 ( 2510 - 5103 \theta ) \theta }
\right)
\right)^{1/3}\ ,
\label{eq:h}
\]
 the average numbers of vertices and skeleton vertices essentially behave as \cite{Ambjorn:2020kpr}\footnote{
 The similar one-parameter family of unconventional continuum limit was originally discussed in a one-matrix model \cite{Ambjorn:2014bga}. 
 } 
\[
\langle V \rangle \sim \frac{1}{\varepsilon^{2}}\ , \ \ \ \langle V_s \rangle \sim \frac{1}{\varepsilon^{2 -2 a /3}}\ . 
\]

In particular, when we set $a = 3$, the continuum limit is characterized by $\langle V \rangle = \infty$ and $\langle V_s \rangle < \infty$. 
Therefore, in this continuum limit, the tree structures become important where the average number of skeleton vertices, or equivalently the average number of index loops, stays finite. 
Since we approach the zero temperature point along with the critical curve given by $c=c_c (\theta)$, 
the fluctuations of spin variables would be divergent, and therefore important in the continuum limit. 
This \textit{is} the continuum limit of branched polymers (BPs) with loops coupled to the critical Ising model, 
which we wish to study in this article.

\subsection{Critical Ising model}
\label{sec:CriticalIsing}

As discussed in Ref.~\cite{Sato:2017ccb}, one can directly take the continuum limit of the matrix integral (\ref{eq:mm}). 
This is because the large-$N$ limit and the unconventional continuum limit given by Eq.~(\ref{eq:alphascaling}) with $a = 3$ commute 
in the sense that one obtains the same loop equation, independent of the order of two limits, 
or in other words, the potential of the matrix model does scale in a non-trivial manner. 

Changing the variables as
\[
\phi_+ = A + \frac{1}{2g} B + \frac{1+c}{2g}\ , \ \ \ \phi_- = - A + \frac{1}{2g} B + \frac{1+c}{2g}\ , 
\label{eq:changeAB}
\]
the integral (\ref{eq:mm}) becomes up to overall constant 
\[
Z_N(g, c,\theta) 
= \int DADB\ e^{\frac{N}{\theta} \text{tr} \left( A^2 B - \widetilde{U}(B) \right) }\ , 
\label{eq:zab}
\]  
where
\[
\widetilde{U}(B) 
=- 
\frac{1}{12g^2}
\left(
B^3 + 6cB^2 + 3 ( 4g^2 + (3c-1)(c+1) )B
\right)\ .  
\label{eq:utilde}
\]
If we choose the following parametrization
\[
g=g_c(\theta) e^{- \frac{1}{2} \lambda_0 \varepsilon^2  }\ , \ \ \ 
c=c_c (\theta)\ , \ \ \ \theta = g_s \varepsilon^3\ , 
\label{eq:continuumlimit}
\]
and if we scale the matrices as 
\[
A = X \varepsilon , \ \ \ B = Y \varepsilon\ , 
\]
the continuum limit of the integral (\ref{eq:zab}) becomes up to overall constant \cite{Sato:2017ccb}
\[
\int DXDY\ e^{\frac{N}{g_s} \text{tr} \left( X^2 Y - V(Y) \right)} =: I_N (\lambda, g_s)
\ ,
\label{eq:I}
\]
where 
\[
V(Y) 
=   \lambda Y - \frac{1}{3} Y^3  
- \frac{\gamma}{2}  g^{1/3}_s Y^2\ , \ \ \ \text{with}\ \ \ \lambda = \lambda_0 + \frac{6}{5^{2/3}} g^{2/3}_s\ , \ \ \ \gamma = 2\times 5^{-1/3}\ .  
\label{eq:vy}
\]
Here $\gamma$ is a numerical constant that can keep truck of the temperature in the original discrete model, because $\gamma$ appears in the expansion, $c_c (\theta) = \frac{\gamma}{4}g^{1/3}_s \varepsilon + \cdots$. 
We wish to stress that through the continuum limit, we have again obtained the matrix model (\ref{eq:I}), but this is the continuum matrix model with the renormalized coupling constants.

If we change the matrix variables as 
\[
X= \frac{1}{2} \left( \Phi_+ - \Phi_- \right)\ , \ \ \ Y=  \frac{1}{2} \left( \Phi_+ + \Phi_- \right) - \frac{\gamma}{4} g^{1/3}_s\ , 
\label{eq:xy}
\]
the integral (\ref{eq:I}) becomes up to overall constant
\[
\int D\Phi_+ D\Phi_-\ e^{ \frac{N}{g_s} \text{tr} \left[ - \frac{\widetilde{\lambda}}{2} \left( \Phi_+ + \Phi_- \right) 
+ \frac{1}{6} \left( \Phi^3_+ + \Phi^3_-  \right)
+ \frac{\gamma}{4} g^{1/3}_s \Phi_+ \Phi_- \right] }\ , 
\label{eq:Ip}
\]
where 
\[
\widetilde{\lambda}  
= \lambda + \frac{3}{16}\gamma^2 g^{2/3}_{s}\ . 
\label{eq:recos}
\]
Note here that one can show that any linear transformation of the Hermitian matrices $X$ and $Y$ to other Hermitian matrices $P$ and $Q$ cannot 
generate a decoupled system such that $X^2 Y + V(Y) = f(P) + g(Q)$ where $f (P)$ and $g(Q)$ are polynomials of $P$ and $Q$, respectively. 

In summary, the continuum matrix model (\ref{eq:I}) or (\ref{eq:Ip}) describes the continuum limit of BPs with loops coupled to the critical Ising model.

\subsection{Absence of Ising coupling} 

Let us turn off the Ising coupling, i.e. $c=0$, and then take the continuum limit 
by choosing the parametrization
\[
g=\widetilde{g_c}(\theta) e^{- \frac{1}{2} \lambda_0 \varepsilon^2  }\ , \ \ \ 
c=0\ , \ \ \ \theta = g_s \varepsilon^3\ , 
\label{eq:continuumlimit2}
\]
where $\widetilde{g_c}(\theta) = g_c (\theta, 0)$ given by   
\[
\widetilde{g}_c (\theta) 
= \left( \frac{1}{4} - \frac{9}{4} \theta^2 + \frac{3 \theta^{2/3} }{4 \times 2^{2/3} q} \left( 2^{1/3} q^2 - 4\theta^{2/3} + 18 \theta^{8/3} \right) \right)^{1/2} \ ; 
\label{eq:tildegc}
\]
\[
q = \left( -1 + 18 \theta^2 -54 \theta^4 + \sqrt{1-4\theta^2} \right)^{1/3}\ .
\label{eq:q}
\]
As before, if one scales the matrices as 
\[
A = X \varepsilon , \ \ \ B = Y \varepsilon\ , 
\label{eq:AB2}
\]
the matrix integrals (\ref{eq:zab}) become up to overall constant
\[
\int DXDY\ e^{\frac{N}{g_s} \text{tr} \left( X^2 Y - \lambda_{\text{cdt}} Y + \frac{1}{3} Y^3 \right)} \ ,
\ \ \ \text{with}\ \ \ 
\lambda_{\text{cdt}} = \lambda_0 + 3 g^{2/3}_s\ . 
\label{eq:pure}
\]
If changing the variables again,
\[
X = \Phi_+ - \Phi_-\ , \ \ \ Y= \Phi_+ + \Phi_-\ , 
\label{eq:XYPhi}
\]
the matrix integral (\ref{eq:pure}) becomes \cite{Sato:2017ccb}
\[
\int D\Phi_+  \ e^{\frac{N}{g_s} \text{tr} \left( -\lambda_{\text{cdt}} \Phi_+ + \frac{4}{3} \Phi^3_+ \right)}
\int D\Phi_-
\ e^{\frac{N}{g_s} \text{tr} \left( - \lambda_{\text{cdt}} \Phi_- + \frac{4}{3} \Phi^3_- \right)}
\ . 
\label{eq:decoupled}
\]
This is the square of the partition function of a continuum theory known as the generalized causal dynamical triangulations (CDT) \cite{Ambjorn:2008jf, Ambjorn:2008gk} 
where $\lambda_{\text{cdt}}$ and $g_s$ are, respectively, 
the renormalized cosmological constant and the renormalized string coupling constant controlling the spatial topology change. 

Setting that $c=0$ from the beginning means that the Hermitian two-matrix integral (\ref{eq:mm}) becomes 
the square of the following Hermitian one-matrix integral
\[
\int D\phi\ e^{- \frac{N}{\theta} \text{tr} \left( \frac{1}{2} \phi^2 - g\phi - \frac{g}{3}\phi^3 \right)}\ .
\label{eq:onemm}
\]
This Hermitian one-matrix model produces the sum over ribbon graphs consisting of vertices of degree one and three.  
The partition function of generalized CDT can be obtained by an unconventional continuum limit 
such that the average number of skeleton vertices stays finite, and the tree structures become important \cite{Ambjorn:2008gk}. 
In this sense, the partition function of generalized CDT can be interpreted as the one of the continuum limit of pure BPs with loops. 
Although the geometries of generalized CDT are not polymer-like but extended,  
they eventually have the same continuum partition function.

Comparing Eq.~(\ref{eq:Ip}) with Eq.~(\ref{eq:decoupled}), 
the effect of divergent fluctuations of spin variables can be expressed by the interaction term proportional to $\gamma$, 
which is absent for the continuum limit of pure BPs with loops.

\subsection{Saddle-point equation in the large-$N$ limit}
Let us perform the Gaussian integral over $X$ in the partition function (\ref{eq:I}). If we introduce the eigenvalues $\tau_i$ of the matrix $Y$, we obtain 
\[
\text{tr} (BA^2) = \frac{1}{2} \sum^{N}_{i=1} \sum^{N}_{j=1} (\tau_i + \tau_j) \bigl|\widetilde{A}_{ij} \bigl|^2\ , 
\label{eq:trBAA}
\]
where 
\[
U^{\dagger} B U = \text{diag} (\tau_1, \tau_2, \cdots , \tau_N)\ , \ \ \ U \in \text{U}(N)\ ; \ \ \ \widetilde{A} = U^{\dagger} A U\ . 
\label{eq:diagonalize}
\]
When the eigenvalue distribution has a compact support around the local minimum of the potential at large $N$ where all the eigenvalues $\tau_i$ are negative, 
one can perform the Gaussian integral, and the partition function becomes up to overall constant
\[
I_N (\lambda, g_s) =
\int \prod_{i} d\tau_i \prod_{j < k} |\tau_j - \tau_k|^2 \prod_{\ell ,m} |\tau_{\ell} + \tau_m|^{-1/2}\ e^{-\frac{N}{g_s} \sum_i V (\tau_i)}\ , 
\label{eq:I2}
\]
where 
\[
V (\tau_i) =   \lambda \tau_i - \frac{1}{3} \tau^3_i  
- \frac{\gamma}{2} g^{1/3}_s \tau^2_i \ . 
\label{eq:Vtau}
\]
In the large-$N$ limit, the saddle-point equation yields
\[
\frac{2g_s}{N} \sum_{j \ne i} \frac{1}{\tau_i - \tau_j} 
= \frac{g_s}{N} \sum_j \frac{1}{\tau_i + \tau_j} + V'(\tau_i)\ . 
\label{eq:saddle}
\]
If we introduce the resolvent, 
\[
\widetilde{w}(z) = \frac{1}{N}
\sum^{N}_{i=1}
\frac{1}{z - \tau_i}
\ , 
\label{eq:resolvent}
\]
multiply the saddle-point equation (\ref{eq:saddle}) by $1/(N(z-\tau_i))$, 
and perform the sum over the label $i$, 
we obtain 
\[
&V'(z) \widetilde{w}(z) + V'(-z) \widetilde{w}(-z) 
- g_s 
\left(
\widetilde{w}^2(z) 
+ \widetilde{w}(z) \widetilde{w}(-z)
+\widetilde{w}^2(-z)
\right) 
- \frac{g_s}{N} \left( \widetilde{w}'(z) + \widetilde{w}'(-z) \right) \notag \\
&\ \ \ = - 2 \int d\tau \rho (\tau) \left(\tau + \gamma g^{1/3}_s \right)\ , 
\label{eq:saddle2}
\]
where $\rho (\tau) = \frac{1}{N} \sum^{N}_{i=1} \delta (\tau - \tau_i)$\footnote{
We have followed the procedure used in Ref.~\cite{Eynard:1992cn}. 
}. 
Differentiating the equation (\ref{eq:saddle2}) with respect to $z$, and taking the large-$N$ limit, we obtain 
\[
\frac{\partial}{\partial z} 
\left( 
\left( 
z^2  - \lambda
\right) \left( \widetilde{w}(z) + \widetilde{w}(-z) \right)
+ \gamma g^{1/3}_s z \left(  \widetilde{w}(z) - \widetilde{w}(-z)  \right)
+ g_s \left( \widetilde{w}^2 (z) + \widetilde{w}^2 (-z) +  \widetilde{w} (z) \widetilde{w} (-z) \right)
\right) = 0\ .
\label{eq:looplargen}
\]

\subsection{Loop equation}

Let us rewrite the partition function (\ref{eq:I}) as 
\[
I_N (\lambda, g_s) 
= \int \prod^{N}_{i=1} d\tau_i\ e^{-\frac{N}{g_s} V_{\text{eff}}(\tau)}\ , 
\label{eq:I3}
\]
where $\tau_i$ are the eigenvalues of $Y$ and 
\[
V_{\text{eff}}(\tau) 
= \sum^{N}_{i=1} 
\left(
V (\tau_i) 
- \frac{g_s}{N} \sum^{N}_{j(\ne i)=1} \ln |\tau_i - \tau_j| 
+ \frac{g_s}{2N} \sum^{N}_{j=1} \ln |\tau_i + \tau_j|
\right)\ , 
\label{eq:vefftau}
\]
and $V(\tau_i)$ is given by Eq.~(\ref{eq:Vtau}).

Assuming that the integrand vanishes at the boundary, we consider the following Dyson-Schwinger equation
\[
0 = \int \prod^{N}_{i=1} d\tau_i\ 
\sum^{N}_{k=1} \frac{\partial}{\partial \tau_k} 
\left(
\frac{1}{z - \tau_k}\
e^{-\frac{N}{g_s} V_{\text{eff}}(\tau)}
\right) 
\ , 
\label{eq:DSmm}
\]
which yields the loop equation
\[
&\left\langle 
\sum^{N}_{k=1} \frac{1}{(z - \tau_k)^2}
\right\rangle 
- \frac{N}{g_s} 
\left\langle 
\sum^{N}_{k=1} \frac{V'(\tau_k)}{z-\tau_k} 
\right\rangle 
+ 
\left\langle 
\sum^{N}_{k=1} \frac{1}{z-\tau_k} 
\sum^{N}_{j(\ne k)=1} \frac{2}{\tau_k - \tau_j} 
\right\rangle \notag \\
& \ \ \ 
- 
\left\langle 
\sum^{N}_{k=1} \frac{1}{z-\tau_k} 
\sum^{N}_{j=1} \frac{1}{\tau_k + \tau_j} 
\right\rangle = 0\ . 
\label{eq:Loop}
\]
Here the expectation value is calculated using $I_N(\lambda, g_s)$. 
Since we have
\[
\left\langle 
\sum^{N}_{k=1} \frac{1}{(z - \tau_k)^2}
\right\rangle 
+ 
\left\langle 
\sum^{N}_{k=1} \frac{1}{z-\tau_k} 
\sum^{N}_{j(\ne k)=1} \frac{2}{\tau_k - \tau_j} 
\right\rangle
= 
\left\langle
\sum^{N}_{k=1} \frac{1}{z-\tau_k}
\sum^{N}_{j=1} \frac{1}{z-\tau_j}
\right\rangle\ , 
\label{eq:identity1}
\]
the loop equation (\ref{eq:Loop}) becomes 
\[
&\left\langle
\sum^{N}_{k=1} \frac{1}{z-\tau_k}
\sum^{N}_{j=1} \frac{1}{z-\tau_j}
\right\rangle
-  \frac{N}{g_s} 
\left\langle 
\sum^{N}_{k=1} \frac{V'(z)}{z-\tau_k} 
\right\rangle 
+ 
\frac{N}{g_s} 
\left\langle 
\sum^{N}_{k=1} \frac{V'(z) - V'(\tau_k)}{z-\tau_k} 
\right\rangle \notag \\
&\ \ \ 
- 
\left\langle 
\sum^{N}_{k=1} \frac{1}{z-\tau_k} 
\sum^{N}_{j=1} \frac{1}{\tau_k + \tau_j} 
\right\rangle = 0\ . 
\label{eq:Loop2}
\]

Introducing the expectation value of the resolvent,  
\[
\widetilde{w}(z) = 
\frac{1}{N}
\left\langle 
\sum^{N}_{k=1} \frac{1}{z-\tau_k} 
\right\rangle
\ , 
\label{eq:vevresolvent}
\] 
we have
\[
\left\langle
\sum^{N}_{k=1} \frac{1}{z-\tau_k}
\sum^{N}_{j=1} \frac{1}{z-\tau_j}
\right\rangle 
= N^2 
\widetilde{w}^2 (z) +  \widetilde{w}(z,z)
\ , 
\label{eq:identity2}
\]
where
\[
\widetilde{w}(z,z) 
:= 
\left\langle 
\sum^{N}_{k=1} \frac{1}{z-\tau_k} 
\sum^{N}_{j=1} \frac{1}{z-\tau_j} 
\right\rangle_{\text{con}}\ . 
\label{eq:connected}
\]
Here $\langle \cdot \rangle_{\text{con}}$ denotes the connected part of the expectation value. 
Using these expressions, the loop equation (\ref{eq:Loop2}) can be recast as
\[
&\widetilde{w}^2 (z) + \frac{1}{N^2} w(z,z) 
-\frac{1}{g_s} V'(z) \widetilde{w}(z) + \frac{1}{g_s N} \sum^{N}_{k=1} 
\left\langle 
 \frac{V'(z) - V'(\tau_k)}{z-\tau_k} 
\right\rangle \notag \\
&\ \ \ 
- \frac{1}{N^2} 
\sum^{N}_{k=1}
\sum^{N}_{j=1}
\left\langle 
\frac{1}{z-\tau_k} 
\frac{1}{\tau_k + \tau_j} 
\right\rangle = 0\ . 
\label{eq:Loop3}
\]
If we change $z$ to $-z$ in the loop equation (\ref{eq:Loop3}) and add it to the original loop equation we obtain
\[
&- \left(
V'(z)\widetilde{w} (z) 
+ V'(-z) \widetilde{w} (-z)
\right) 
+ g_s \left(
\widetilde{w}^2 (z) 
+ \widetilde{w}^2 (-z) 
+ \widetilde{w} (z)\widetilde{w} (-z)
\right) \notag \\
&\ \ \ 
+ \frac{g_s}{N^2} 
\left( 
\widetilde{w} (z,z) 
+ \widetilde{w} (-z,-z) 
+ \widetilde{w} (z,-z)
\right) 
= \frac{2}{N} \sum^{N}_{k=1} \left\langle \tau_k + \gamma g^{1/3}_s \right\rangle\ . 
\label{eq:Loop4}
\]
Differentiating the equation above with respect to $z$ yields
\[
&\frac{\partial}{\partial z} 
\biggl[
(z^2-\lambda) \left( \widetilde{w} (z) + \widetilde{w} (-z) \right) + \gamma g^{1/3}_s z \left( \widetilde{w} (z) - \widetilde{w} (-z) \right) 
+ g_s \left(
\widetilde{w}^2 (z) 
+ \widetilde{w}^2 (-z) 
+ \widetilde{w} (z)\widetilde{w} (-z)
\right) \notag \\
&\ \ \ 
+ \frac{g_s}{N^2} 
\left( 
\widetilde{w} (z,z) 
+ \widetilde{w} (-z,-z) 
+ \widetilde{w} (z,-z)
\right) 
\biggl] 
= 0\ . 
\label{eq:Loop5}
\]

As explained in the next section, one can construct a string field theory that reproduces the loop equation (\ref{eq:Loop5}) as the Dyson{\textendash}Schwinger equation.

\section{String field theory}
\label{sec:sft}

There exist maps between branched polymers (BPs) and discrete surfaces based on causal dynamical triangulations (CDT) where spatial topology change is not allowed to occur (see, e.g. \cite{Durhuus:2009sm}). 
Such maps can be generalized to a map between BPs with loops and discrete generalized surfaces where the spatial topology can change \cite{Ambjorn:2013csx}. 

 Having this in mind, it may be possible to construct a string field theory that describes the continuum limit of BPs with loops coupled to the critical Ising model. 
 In fact, a string field theory for generalized CDT has been constructed \cite{Ambjorn:2008ta}, and the generalized CDT can be interpreted as the continuum limit of pure BPs with loops.

Let us define an operator $\psi^{\dagger} (\ell)$ that creates (marked) closed string of length $\ell$.  
Similarly, an operator $\psi (\ell)$ annihilates (marked) closed string of length $\ell$.   
The operators satisfy the commutation relation
\[
[ \psi (\ell), \psi^{\dagger} (\ell') ] = \ell \delta (\ell - \ell')\ , 
\label{eq:commutationrelation}
\]
and others are zero. 
The vacuum state $| \text{vac} \rangle$ is defined by 
\[
\psi (\ell) | \text{vac} \rangle = 0 \ \ \ \Leftrightarrow \ \ \   \langle \text{vac} | \psi^{\dagger} (\ell)  = 0\ . 
\label{eq:vacuum}
\]

We conjecture the following Hamiltonian 
\[
\mathcal{H} 
&= \int^{\infty}_{0} \frac{d\ell}{\ell}\ \psi^{\dagger} (\ell) \left(- \ell \frac{\partial^2}{\partial \ell^2} + \lambda \ell \right) \psi (\ell) \ \ \ &(\text{Free}) \notag \\
&-g_s \int^{\infty}_{0} d\ell_1 \int^{\infty}_{0} d \ell_2\ \psi^{\dagger} (\ell_1)  \psi^{\dagger} (\ell_2)  \psi (\ell_1 + \ell_2) \ \ \  &(\text{Splitting interaction}) \notag \\
&-\alpha g_s \int^{\infty}_{0} d\ell_1 \int^{\infty}_{0} d \ell_2\ \psi^{\dagger} (\ell_1 + \ell_2)  \psi (\ell_1)  \psi (\ell_2)  \ \ \  &(\text{Joining interaction}) \notag \\
&-\beta g_s \int^{\infty}_{0} d\ell_1 \int^{\infty}_{0} d \ell_2\ \psi^{\dagger} (\ell_1 + \ell_2)  \psi^{\dagger} (\ell_2)  \psi (\ell_1) \ \ \  &(\text{Effective interaction}) \notag \\
&- \gamma g^{1/3}_s \int^{\infty}_{0} \frac{d\ell}{\ell}\ \psi^{\dagger} (\ell) \left( - \ell \frac{\partial}{\partial \ell}  \right) \psi (\ell)\ \ \  &(\text{Hinge}) \notag \\
&-  \int^{\infty}_{0} \frac{d\ell}{\ell}\ \delta (\ell) \psi (\ell)\ , \ \ \  &(\text{Tadpole})
\label{eq:calH}
\]
where $\alpha$, $\beta$ and $\gamma$ are dimensionless constants, and if we denote the dimension of length $\ell$ as $[\ell] = L$, then 
\[
[\psi (\ell)] = [\psi^{\dagger} (\ell)]   = L^0\ , \ \ \ [\lambda] = L^{-2}\ , \ \ \ [g_s] = L^{-3}\ , \ \ \ [\mathcal{H}] = L^{-1}\ . 
\label{eq:dimensions}
\]

String field theories for non-critical stings have been proposed by Ref.~\cite{Ishibashi:1993nq, Ishibashi:1993nqz}. 
In particular, to write down the Hamiltonian (\ref{eq:calH}), we have referred to Refs.~\cite{Ishibashi:1993nqz, Ambjorn:2008ta, Fuji:2011ce}. 
The basic four parts in Eq.~(\ref{eq:calH}), i.e. the free, splitting-interaction, joining-interaction and tadpole parts, are the building blocks of the string field theory for generalized CDT \cite{Ambjorn:2008ta}. 

The term similar to the effective-interaction part in Eq.~(\ref{eq:calH}) has been originally introduced in Ref.~\cite{Ishibashi:1993nqz}, and also discussed in Ref.~\cite{Fuji:2011ce}. 
This non-local term is necessary for describing the non-trivial Jacobian factor $\prod_{\ell ,m} |\tau_{\ell} + \tau_m|^{-1/2}$ in the continuum two-matrix model (\ref{eq:I2}), 
which appears after performing the Gaussian integral.  

The hinge-part in Eq.~(\ref{eq:calH}) is introduced to take in the effects from the $\Phi_+ \Phi_-$ coupling in the continuum two-matrix model (\ref{eq:Ip}), 
which would be a hinge to capture the quantum critical behavior of the Ising model on BPs with loops.

\subsection{Sphere case}

Let us set $\alpha = 0$. 
The loop function with the disk topology, 
\[
w (\ell) = \lim_{t \to \infty} \langle \text{vac} | e^{-t \mathcal{H} } \psi^{\dagger} (\ell) | \text{vac} \rangle\ , 
\label{eq:diskamplitude}
\]
satisfies the Dyson{\textendash}Schwinger equation that can be interpreted as the Wheeler{\textendash}DeWitt equation 
\[
0 = - \lim_{t \to \infty} \frac{\partial}{\partial t} \langle \text{vac} | e^{-t \mathcal{H}} \psi^{\dagger} (\ell) | \text{vac} \rangle 
= \lim_{t \to \infty}  \langle \text{vac} | e^{-t \mathcal{H} } [\mathcal{H} , \psi^{\dagger} (\ell)] | \text{vac} \rangle\ .
\label{eq:SFTSD}
\]
In the last line we have used $\mathcal{H} | \text{vac} \rangle = 0$. 

We calculate Eq.~(\ref{eq:SFTSD}) term by term. 
\begin{enumerate}
\item The free part: 
\[
&\left[ 
\int^{\infty}_{0} \frac{d\ell'}{\ell'}\ \psi^{\dagger} (\ell') \left(- \ell' \frac{\partial^2}{\partial \ell'^2} + \lambda \ell' \right) \psi (\ell'), \psi^{\dagger} (\ell)
\right]\notag \\
&\ \ \ = \int^{\infty}_{0} \frac{d\ell'}{\ell'}\ \psi^{\dagger} (\ell') \left(- \ell' \frac{\partial^2}{\partial \ell'^2} + \lambda \ell' \right) \left( \ell \delta (\ell - \ell') \right) \notag \\
&\ \ \ = \int^{\infty}_{0} \frac{d\ell'}{\ell'}\ \left( \ell \delta (\ell - \ell') \right) \left(- \ell' \frac{\partial^2}{\partial \ell'^2} + \lambda \ell' \right) \psi^{\dagger} (\ell') \notag \\
&\ \ \ = \ell \left(- \frac{\partial^2}{\partial \ell^2} + \lambda  \right) \psi^{\dagger} (\ell)\ . 
\label{eq:free}
\]
\item The splitting-interaction part:
\[
&\left[ 
-g_s \int^{\infty}_{0} d\ell_1 \int^{\infty}_{0} d \ell_2\ \psi^{\dagger} (\ell_1)  \psi^{\dagger} (\ell_2)  \psi (\ell_1 + \ell_2), \psi^{\dagger} (\ell)
\right] \notag \\
&\ \ \ 
= -g_s \int^{\infty}_{0} d\ell_1 \int^{\infty}_{0} d \ell_2\ \psi^{\dagger} (\ell_1)  \psi^{\dagger} (\ell_2)\ \ell \delta (\ell_1 + \ell_2 - \ell) \notag \\
&\ \ \ 
= -g_s \int^{\ell}_{0} d\ell_1\ \psi^{\dagger} (\ell_1)  \psi^{\dagger} (\ell - \ell_1)\ \ell\ . 
\label{eq:splitting}
\]
\item The effective-interaction part:
\[
&\left[
- \beta g_s \int^{\infty}_{0} d\ell_1 \int^{\infty}_{0} d \ell_2\ \psi^{\dagger} (\ell_1 + \ell_2)  \psi^{\dagger} (\ell_2)  \psi (\ell_1), \psi^{\dagger} (\ell)
\right]\notag \\
&\ \ \ 
= - \beta g_s \int^{\infty}_{0} d\ell_1 \int^{\infty}_{0} d \ell_2\ \psi^{\dagger} (\ell_1 + \ell_2)  \psi^{\dagger} (\ell_2)\ \ell_1 \delta (\ell_1 - \ell)\notag \\
&\ \ \ 
= - \beta g_s \ell \int^{\infty}_{0} d \ell_2\ \psi^{\dagger} (\ell + \ell_2)  \psi^{\dagger} (\ell_2) \ . 
\label{eq:effective}
\]
\item The hinge part:
\[
&\left[
-\gamma g^{1/3}_s \int^{\infty}_{0} 
\frac{d\ell'}{\ell'}\ 
\psi^{\dagger} (\ell') 
\left( - \ell' \frac{\partial}{\partial \ell'} \right) 
\psi (\ell'), 
\psi^{\dagger} (\ell)
\right] \notag \\
&\ \ \ 
= -\gamma g^{1/3}_s \int^{\infty}_{0}d\ell' \ \frac{\partial}{\partial \ell'} \psi^{\dagger} (\ell')\ \ell' \delta (\ell' - \ell) \notag \\
&\ \ \ 
= -\gamma g^{1/3}_s \frac{\partial}{\partial \ell} \psi^{\dagger} (\ell)\ \ell\ .  
\label{eq:hinge}
\]
\item The tadpole part:
\[
&\left[
-  \int^{\infty}_{0} \frac{d\ell'}{\ell'}\ \delta (\ell') \psi (\ell'), \psi^{\dagger} (\ell)
\right]\notag \\
&\ \ \ 
= - \int^{\infty}_{0} \frac{d\ell'}{\ell'}\ \delta (\ell')\ \ell \delta (\ell - \ell') \notag \\
&\ \ \ 
= - \delta (\ell)\ . 
\label{eq:tadpole}
\]
\end{enumerate}
As a result, we obtain 
\[
[\mathcal{H} , \psi^{\dagger} (\ell)] 
&= \ell \left(- \frac{\partial^2}{\partial \ell^2} + \lambda - \gamma g^{1/3}_s \frac{\partial}{\partial \ell}    \right) \psi^{\dagger} (\ell) 
- g_s \ell \int^{\ell}_{0} d\ell_1\ \psi^{\dagger} (\ell_1)  \psi^{\dagger} (\ell - \ell_1) \notag \\
&\ \ \   
- \beta g_s \ell \int^{\infty}_{0} d \ell_2\ \psi^{\dagger} (\ell + \ell_2)  \psi^{\dagger} (\ell_2)
- \delta (\ell)\ .
\label{eq:calHpsidagger}
\]
Since we have turned off the joining interaction by setting $\alpha = 0$, we notice the factorization
\[
\lim_{t \to \infty} \langle \text{vac} | e^{-t \mathcal{H} } \psi^{\dagger} (\ell_1) \psi^{\dagger} (\ell_2) | \text{vac} \rangle 
= \lim_{t \to \infty} \langle \text{vac} | e^{-t \mathcal{H} } \psi^{\dagger} (\ell_1)  | \text{vac} \rangle \lim_{t \to \infty} \langle \text{vac} | e^{-t \mathcal{H} } \psi^{\dagger} (\ell_2) | \text{vac} \rangle\ . 
\label{eq:factorization}
\]
In terms of all the discussions above, we obtain 
\[
 \ell \left(- \frac{\partial^2}{\partial \ell^2} + \lambda - \gamma g^{1/3}_s \frac{\partial}{\partial \ell}    \right) w (\ell) 
 - g_s \ell \left( w \ast w \right) (\ell) 
- \beta g_s \ell \int^{\infty}_{0} d \ell_2\ w (\ell + \ell_2)  w (\ell_2)
 = \delta (\ell)\ , 
 \label{eq:wdw}
\]
where the convolution of $w$ is given by
\[
\left( w \ast w \right) (\ell) 
= \int^{\ell}_{0} d\ell_1\ w (\ell_1)  w (\ell - \ell_1)\ . 
\label{eq:convolution}
\]

Let us define the Laplace transform of $w(\ell)$
\[
\widetilde{w} (z) = \mathcal{L} [ w(\ell) ] = \int^{\infty}_{0}d\ell\ e^{-z\ell} w(\ell)\ . 
\label{eq:Lw}
\]
Using Eq.~(\ref{eq:Lw}), the Laplace transform of Eq.~(\ref{eq:wdw}) becomes
\[
\frac{\partial}{\partial z} 
\left( 
V'(z) \widetilde{w}(z) 
+ g_s \widetilde{w}^2 (z) 
+ \beta g_s \mathcal{L} \left[ \int^{\infty}_{0} d \ell_2\ w (\ell + \ell_2)  w (\ell_2) \right]
\right) = 1\ , 
\label{eq:wdw2}
\]
where 
\[
V'(z) = z^2 - \lambda + \gamma g^{1/3}_s z \ . 
\label{eq:Vpz}
\]
Here the term $\mathcal{L} \left[ \int^{\infty}_{0} d \ell_2\ w (\ell + \ell_2)  w (\ell_2) \right]$ would include a divergence, 
which seems to be regularized by subtracting Eq.~(\ref{eq:wdw2}) by the one with the replacement $z \to -z$ \cite{Ishibashi:1993nqz, Fuji:2011ce}
\[
\frac{\partial}{\partial z} 
\left( 
\left( 
z^2  - \lambda
\right) \left( \widetilde{w}(z) + \widetilde{w}(-z) \right)
+ \gamma g^{1/3}_s z \left(  \widetilde{w}(z) - \widetilde{w}(-z)  \right)
+ g_s \left( \widetilde{w}^2 (z) + \widetilde{w}^2 (-z) + \beta \widetilde{w} (z) \widetilde{w} (-z) \right)
\right) = 0\ .
\label{eq:sde}
\]
If we set $\beta = 1$, this equation reproduces the saddle-point equation in the continuum two-matrix model (\ref{eq:looplargen}).

\subsection{General amplitudes}
We introduce the generating functional $Z(J)$ for the loop functions
\[
Z(J) 
= \lim_{t \to \infty} 
\langle \text{vac} | 
e^{-t \mathcal{H}} e^{\int^{\infty}_{0} d\ell\ J (\ell) \psi^{\dagger} (\ell) }
| \text{vac} \rangle\ .
\label{eq:ZJ}
\]
The Dyson{\textendash}Schwinger equation yields
\[
0 = - \lim_{t \to \infty} \frac{\partial}{\partial t}
\langle \text{vac} | 
e^{-t \mathcal{H}} e^{\int^{\infty}_{0} d\ell\ J (\ell) \psi^{\dagger} (\ell) }
| \text{vac} \rangle\ ,
\label{eq:SFTSD2}
\]
which can be recast as
\[
0 = \int^{\infty}_0d\ell\ 
J (\ell) T (\ell) 
\ Z(J)\ , 
\label{eq:sd_full}
\]
where
\[
T (\ell) 
&= \left( - \ell \frac{\partial^2}{\partial \ell^2} + \lambda \ell \right) \frac{\delta}{\delta J (\ell)} \ \ \ &(\text{Free}) \notag \\
& \ \ \ - g_s \ell \int^{\ell}_{0}d\ell'\ \frac{\delta^2}{\delta J (\ell') \delta J (\ell - \ell')}  \ \ \  &(\text{Splitting interaction}) \notag \\
& \ \ \ - \alpha g_s \ell \int^{\infty}_{0}d\ell'\ \ell' J (\ell') \frac{\delta}{\delta J (\ell + \ell')} \ \ \  &(\text{Joining interaction})  \notag \\
& \ \ \ - \beta g_s \ell \int^{\infty}_{0}d\ell'\ \frac{\delta^2}{\delta J (\ell + \ell') \delta J (\ell') } \ \ \  &(\text{Effective interaction}) \notag \\
& \ \ \ - \gamma g^{1/3}_s  \ell \frac{\partial}{\partial \ell}  \frac{\delta}{\delta J (\ell)} \ \ \  &(\text{Hinge})  \notag \\
& \ \ \ - \delta (\ell)\ . \ \ \  &(\text{Tadpole}) 
\label{eq:Tl}
\]

Introducing 
\[
F(J) = \log Z(J)\ , 
\label{eq:FJ}
\]
one can rewrite Eq.~(\ref{eq:sd_full}) into the following form
\[
0 &= \int^{\infty}_{0}d\ell\ J (\ell) 
\biggl{[}
\left( - \ell \frac{\partial^2}{\partial \ell^2} + \lambda \ell \right) \frac{\delta F(J)}{\delta J (\ell)} \notag \\
& \ \ \ - g_s \ell \int^{\ell}_{0}d\ell'\ \frac{\delta^2 F(J) }{\delta J (\ell') \delta J (\ell - \ell')} 
- g_s \ell \int^{\ell}_{0}d\ell'\ \frac{\delta F(J) }{\delta J (\ell') } \frac{ \delta F(J) }{ \delta J (\ell - \ell') } \notag \\
& \ \ \ - \alpha g_s \ell \int^{\infty}_{0}d\ell'\ \ell' J (\ell') \frac{\delta F(J)}{\delta J (\ell + \ell')} \notag \\
& \ \ \ - \beta g_s \ell \int^{\infty}_{0}d\ell'\ \frac{\delta^2 F(J)}{\delta J (\ell + \ell') \delta J (\ell') }
- \beta g_s \ell \int^{\infty}_{0}d\ell'\ \frac{\delta F(J)}{\delta J (\ell + \ell')  }\frac{\delta F(J)}{\delta J (\ell')} \notag \\
& \ \ \ - \gamma g^{1/3}_s \ell \frac{\partial}{\partial \ell}  \frac{\delta F (J)}{\delta J (\ell)} \notag \\
& \ \ \ - \delta (\ell)
\biggl{]} 
\ . 
\label{eq:SDeq_connected}
\]
To derive Eq,~(\ref{eq:SDeq_connected}), we have used
\[
\frac{\delta^2 F(J)}{\delta J (\ell') \delta J (\ell - \ell') } 
+ \frac{ \delta F(J) }{\delta J (\ell')} \frac{ \delta F(J) }{\delta J (\ell - \ell')}
= \frac{1}{Z(J)}
\frac{\delta^2 Z(J)}{\delta J (\ell') \delta J (\ell - \ell') }\ .
\label{eq:id}
\]

We define the multi-loop function 
\[
w(\ell_1, \cdots , \ell_n) 
= \alpha^{1-n} \frac{\delta^{n} F(J)}{\delta J (\ell_1) \cdots \delta J (\ell_n)} \biggl{|}_{J = 0} \ ,
\label{eq:multiloop}
\]
and its Laplace transform
\[
\widetilde{w} (z_1, \cdots , z_n ) 
= \int^{\infty}_0 d\ell_1 \cdots \int^{\infty}_0 d\ell_n\ 
e^{-z_1 \ell_1 - \cdots - z_n \ell_n}
w(\ell_1, \cdots , \ell_n)\ . 
\label{eq:Lapmultiloop}
\]
Differentiating Eq.~(\ref{eq:SDeq_connected}) with respect to $ J (\ell)$, and then setting $J (\ell) = 0$, 
one obtains
\[
& \ell \left(- \frac{\partial^2}{\partial \ell^2} + \lambda - \gamma g^{1/3}_s \frac{\partial}{\partial \ell}    \right) w (\ell) 
 - g_s \ell \left( w \ast w \right) (\ell) 
- \beta g_s \ell \int^{\infty}_{0} d \ell'\ w (\ell + \ell')  w (\ell') \notag \\
 &\ \ \ - \alpha g_s \ell \int^{\ell}_0 d\ell'\ w(\ell', \ell-\ell') 
 - \alpha \beta g_s \ell \int^{\infty}_0 d\ell'\ w(\ell+\ell', \ell')
 = \delta (\ell)\ . 
 \label{eq:diffJ} 
\]
The Laplace transform of Eq.~(\ref{eq:diffJ}) yields
\[
&\frac{\partial}{\partial z}
\biggl\{
\left( z^2 - \lambda + \gamma g^{1/3}_s z \right) \widetilde{w}(z) 
+ g_s \left( \widetilde{w}^2(z) + \alpha \widetilde{w}(z,z) \right) \notag \\
& \ \ \ 
+ \beta g_s \mathcal{L} \left[ \int^{\infty}_{0} d \ell'\ w (\ell + \ell')  w (\ell') + \alpha \int^{\infty}_0 d\ell'\ w(\ell+\ell', \ell')   \right]
\biggl\}
= 1\ . 
\label{eq:LapdiffJ}
\]
Changing $z$ to $-z$ in the equation above and subtracting it from the original equation yields
\[
&\frac{\partial}{\partial z} 
\biggl[
(z^2-\lambda) \left( \widetilde{w} (z) + \widetilde{w} (-z) \right) + \gamma g^{1/3}_s z \left( \widetilde{w} (z) - \widetilde{w} (-z) \right) 
+ g_s \left(
\widetilde{w}^2 (z) 
+ \widetilde{w}^2 (-z) 
+ \beta \widetilde{w} (z)\widetilde{w} (-z)
\right) \notag \\
&\ \ \ 
+ g_s \alpha 
\left( 
\widetilde{w} (z,z) 
+ \widetilde{w} (-z,-z) 
+ \beta \widetilde{w} (z,-z)
\right) 
\biggl] 
= 0\ . 
\]
This equation indeed coincides with the loop equation (\ref{eq:Loop5}) obtained in the continuum two-matrix model when we set $\alpha = \frac{1}{N^2}$ and $\beta = 1$.

\section{Non-perturbative formulation}
\label{sec:nonperturbative}

The continuum limit of branched polymers (BPs) with loops coupled to the critical Ising model can be described by the continuum two-matrix model (\ref{eq:I}), or equivalently (\ref{eq:Ip}). 
The size of matrices, $N$, serves as a weight factor to discriminate different topologies. 
If we wish to implement the sum over all genera in an equal manner, we can simply set $N=1$, and define the partition function in such a way that it converges. 
In this section, we investigate such a non-perturbative partition function.  

Let us set $N=1$ in the continuum two-matrix model (\ref{eq:I}), which yields
\[
I_{N=1} (\lambda, g_s) 
= \int d\widetilde{x}d\widetilde{y}\  e^{\frac{1}{g_s} \left( \widetilde{y}  \widetilde{x}^2 - \lambda \widetilde{y} + \frac{1}{3} \widetilde{y}^3 + \frac{\gamma}{2} g^{1/3}_s \widetilde{y}^2 \right)}\ . 
\label{eq:IN1}
\]
Changing the dimension-full variables $\widetilde{x}$, $\widetilde{y}$ to the dimensionless variables $x$ and $y$,
\[
\widetilde{x} = g^{1/3}_s x\ , \ \ \ \widetilde{y} = g^{1/3}_s y\ , 
\label{eq:tildexy}
\]
Eq.~(\ref{eq:IN1}) becomes up to overall dimensional constants
\[
Z(t) 
= \iint_D dxdy\ 
e^{yx^2 - t y + \frac{1}{3} y^3 + \frac{\gamma}{2} y^2 }\ , 
\label{eq:zt}
\]
where $t$ is the dimensionless cosmological constant given by
$
t = \frac{\lambda}{g^{2/3}_s}
$. 
Here we need to choose an integration domain $D$ in such a way that the integral converges.

\subsection{Integration domains}
\label{sec:integrationdomain}

We first impose the condition that one can perform the Gaussian integral over $x$. 
If we parametrize $x$ and $y$ as 
\[
x = s e^{i\theta_x}\ , \ \ \ y=r e^{i\theta_y}\ , 
\label{eq:sr}
\]
where
\[
s \in \mathbb{R}\ , \ \ \ \theta_x \in \left(- \frac{\pi}{2}, \frac{\pi}{2} \right]\ ; \ \ \  r \in \mathbb{R}_+\ , \ \ \ \theta_y \in (0, 2\pi )\ ,
\label{eq:domainsr}
\]
we have
\[
x^2 y =  s^2 r e^{i (\theta_y + 2 \theta_x)}\ . 
\label{eq:x2y}
\]
In order to perform the Gaussian integral, we need
\[
\mathrm{Re} \left( e^{i (\theta_y + 2 \theta_x)} \right) < 0 \ \ \ 
\Leftrightarrow \ \ \ 
 \frac{\pi}{2} < \theta_y + 2 \theta_x < \frac{3}{2}\pi  \ \ \ 
\Leftrightarrow \ \ \ 
\frac{\pi}{2} - 2 \theta_x < \theta_y < \frac{3}{2}\pi - 2 \theta_x\ .  
\label{eq:cond0}
\]

After performing the Gaussian integral, we impose that the integrand converges when $r$ goes to infinity. 
Since for $r \gg 1$ we have 
\[
- t y + \frac{1}{3} y^3 + \frac{\gamma}{2} y^2  \sim  \frac{1}{3} y^3 =  \frac{1}{3} e^{3i\theta_y} r^3 \ , 
\label{eq:asymptotic1}
\]
the condition we need is 
\[
\mathrm{Re} \left( e^{3i\theta_y}  \right) = \cos (3\theta_y) < 0\ , 
\label{eq:asymptotic2}
\]
which yields
\[
 \frac{\pi}{2} < 3 \theta_y < \frac{3}{2}\pi \ \ \  &\Leftrightarrow \ \ \ (\mathrm{I}):\ \frac{\pi}{6} <  \theta_y < \frac{\pi}{2}\label{eq:cond1} \\
 \frac{\pi}{2} < 3 \theta_y - 2\pi < \frac{3}{2}\pi \ \ \  &\Leftrightarrow \ \ \ (\mathrm{I\hspace{-1.2pt}I}):\  \frac{5}{6}\pi  <  \theta_y < \frac{7}{6}\pi \label{eq:cond2} \\
 \frac{\pi}{2} < 3 \theta_y - 4\pi < \frac{3}{2}\pi \ \ \  &\Leftrightarrow \ \ \ (\mathrm{I\hspace{-1.2pt}I\hspace{-1.2pt}I}):\  \frac{3}{2}\pi <  \theta_y <  \frac{11}{6}\pi \label{eq:cond3} 
\ . 
\]

After performing the Gaussian integral, the factor
\[
\sqrt{ \frac{\pi}{ e^{i ( \theta_y + 2\theta_x + \pi )} r } } 
= e^{-i \left( \frac{\theta_y}{2} + \theta_x + \frac{\pi}{2} \right) } \sqrt{ \frac{\pi}{r} }\ , 
\]
appears. Therefore, we need a branch cut starting from the origin. 
Let us place the branch cut on the positive real axis in the complex $y$ plane to make the integrand single-valued. 

\begin{figure}[h]
\centering
\includegraphics[width=3.0in]{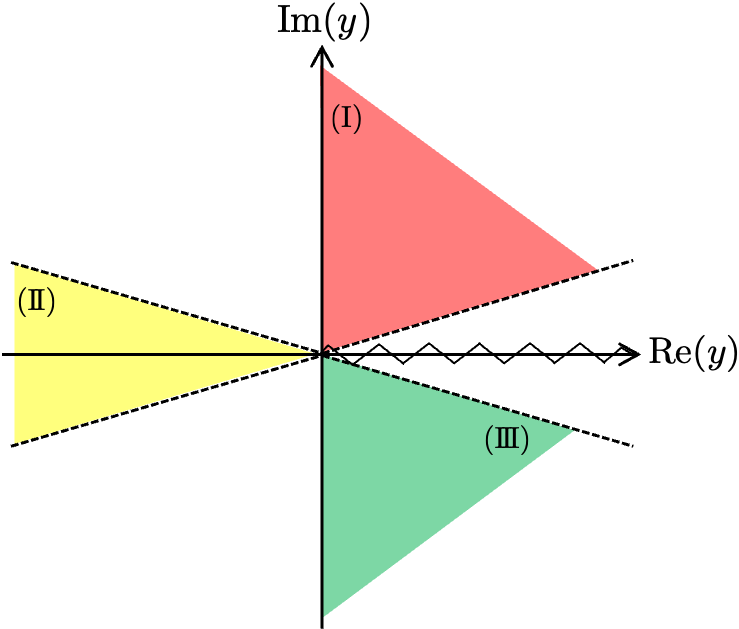}
\caption{Domains of convergence in the complex $y$ plane: The branch cut is placed on the positive real axis.}
\label{fig:asymptotic}
\end{figure}

Hereafter we choose the integration path, denoted by $C$, to run from the asymptotic region ($\mathrm{I\hspace{-1.2pt}I}$) to the asymptotic region $(\mathrm{I})$ (see Figure \ref{fig:asymptotic}). 
This is possible if we choose, for instance, $\theta_x$ to be $\theta_x = \frac{\pi}{6}$. The integration path in the complex-$x$ plane is denoted as $C_x$. 
We then have
\[
Z(t) 
&=\iint_D dxdy\ 
e^{y x^2 - t y + \frac{1}{3} y^3 + \frac{\gamma}{2} y^2 } \notag \\
&= \int_{C_x} dx \int_C  dy\ 
e^{y x^2 - t y + \frac{1}{3} y^3 + \frac{\gamma}{2} y^2 } \notag \\
&=   \int_{C} dy \sqrt{ \frac{\pi}{-y}}\ 
e^{ - t y + \frac{1}{3} y^3 + \frac{\gamma}{2} y^2 }\ . 
\label{eq:Zt2}
\]

\subsection{The third-order differential equation}
\label{sec:thirdorderdiff}

We wish to derive the differential equation that the integral (\ref{eq:zt}) satisfies. 
Let us calculate the Dyson{\textendash}Schwinger equation
\[
0 
&= \iint_D  dxdy \ 
\frac{\partial}{\partial y}
\left(
e^{y x^2 - t y + \frac{1}{3} y^3 + \frac{\gamma}{2} y^2}
\right) \notag \\
&= \iint_D  dxdy \ 
\left( -t  + y^2 + \gamma y \right)
e^{ y x^2 - t y + \frac{1}{3} y^3 + \frac{\gamma}{2} y^2 }
+ \int_{C}  dy\ 
e^{- t y + \frac{1}{3} y^3 + \frac{\gamma}{2} y^2}\ 
\frac{d}{d y} \left( \int_{C_x} dx\ e^{x^2 y} \right) \notag \\
&= - t Z (t) +Z''(t) - \gamma Z'(t)
+ \int_{C} dy\ 
e^{- t y + \frac{1}{3} y^3 + \frac{\gamma}{2} y^2}\ 
\frac{d}{d y}  \sqrt{\frac{\pi}{-y}}  \notag \\
& =- t Z (t) +Z''(t) - \gamma Z'(t)
- \frac{1}{2} \int_C dy\ 
e^{- t y + \frac{1}{3} y^3 + \frac{\gamma}{2} y^2}\ 
\frac{1}{y} \sqrt{\frac{\pi}{ - y}} \notag \\
&= - t Z (t) +Z''(t) - \gamma Z'(t)
- \frac{1}{2} \iint_{D}  dxdy \ 
\frac{1}{y}\
e^{y x^2 - t y + \frac{1}{3} y^3 + \frac{\gamma}{2} y^2 }
\ .
\label{eq:SDZ}
\] 
Differentiating the equation above with respect to $t$ yields  
the third-order linear differential equation 
\[
\left( tZ (t) - Z'' (t) + \gamma Z' (t) \right)' = \frac{1}{2} Z (t)
 \ \ \ \Leftrightarrow\ \ \ 
 Z'''(t) - \gamma Z''(t) - tZ'(t) - \frac{1}{2}Z (t) = 0\ .
 \label{eq:diffe}
\]
In fact, the $\gamma=0$ case is called the Pairy equation: 
``Pairy'' was named after $(\text{Product}+\text{Airy})/2$ \cite{Nakano:1999}, 
and its solution is given by the product of two Airy functions. 
This is natural because the partition function of the continuum limit of pure BPs with loops is known to be Airy functions \cite{Nishigaki:1990sk, Anderson:1991ku, Jurkiewicz:1996yd}.

If we introduce $u(t) = - F'(t)$ such that $Z(t) = e^{-F(t)}$, Eq.~(\ref{eq:diffe}) can be recast as 
\[
u''(t) + (3u(t)-\gamma)u'(t) +u^3(t) - \gamma u^2(t) -tu(t) - \frac{1}{2} = 0\ . 
\label{eq:u}
\]

Let us check the validity of the differential equation (\ref{eq:diffe}). 
We insert Eq.~(\ref{eq:zt}) into the left-hand side of Eq.~(\ref{eq:diffe}), which yields
\[
Z'''(t) - \gamma Z''(t) - tZ'(t) 
&=  \iint_D  dxdy\ 
\left( 
- y^3 - \gamma y^2 + t y
\right)
e^{y x^2 - t y + \frac{1}{3} y^3 + \frac{\gamma}{2} y^2  }  \notag \\
&= - \iint_D  dxdy\ e^{yx^2} y \frac{d}{d y} \left( e^{ -t y + \frac{1}{3} y^3 + \frac{\gamma}{2} y^2} \right) \notag \\
&=  \iint_D  dxdy\  \frac{\partial}{\partial y} \left( y e^{yx^2} \right) e^{ -t y + \frac{1}{3} y^3 + \frac{\gamma}{2} y^2} \notag \\
&= Z(t) +  \int_{C} dy\ y \frac{d}{d y} \left(  \int_{C_x} dx\ e^{yx^2} \right) e^{ - t y + \frac{1}{3} y^3 + \frac{\gamma}{2} y^2} \notag \\
&= Z(t) + \int_C dy\ y \frac{d}{d y}   \sqrt{ \frac{\pi}{- y} } e^{ - t y + \frac{1}{3} y^3 + \frac{\gamma}{2} y^2} \notag \\
&= Z(t) - \frac{1}{2} Z(t) \notag \\ 
&= \frac{1}{2} Z(t)\ . 
\label{eq:check}
\]
Therefore, we have confirmed that $Z(t)$ satisfies the differential equation (\ref{eq:diffe}).

\subsection{Free energy}
\label{eq:freeenergy}

We evaluate the perturbative part of free energy based on the saddle-point approximation.  
Let us rewrite the partition function (\ref{eq:Zt2}) as 
\[
Z(t) = 
\int_C dy\ \sqrt{ \frac{\pi}{- y} }\ e^{-t \widetilde{W}(y)}\ , \ \ \ \text{with}\ \ \ 
\widetilde{W}(y) = y - \frac{1}{3t} y^3 - \frac{\gamma}{2t}y^2\ . 
\label{eq:Zt3}
\]
The saddle-point equation at large-$t$ yields
\[ 
\frac{d}{dy} \widetilde{W}(y) = 0 \ \ \ \Leftrightarrow \ \ \ 1 - \frac{y^2}{t} - \frac{\gamma}{t}y = 0\ .
\label{eq:saddlefree}
\]
The solution is
\[
y_* = - \frac{1}{2} \left( \gamma + \sqrt{ \gamma^2  + 4t} \right)\ . 
\label{eq:yast}
\]
Expanding $\widetilde{W}(y)$ around $y_*$, one obtains
\[
\widetilde{W}(y) 
&= \widetilde{W}(y_*) 
+ \widetilde{W}'(y_*) \left( y - y_* \right) 
+ \frac{1}{2!} \widetilde{W}''(y_*) \left( y - y_* \right)^2 + \frac{1}{3!} \widetilde{W}'''(y_*) \left( y - y_* \right)^3 \notag \\
&= \widetilde{W}(y_*) + \frac{\sqrt{\gamma^2 + 4t}}{2t} \left( y - y_* \right)^2 - \frac{1}{3t} \left( y - y_* \right)^3\ , 
\label{eq:expandwtilde1}
\] 
where $\frac{\sqrt{\gamma^2 + 4t}}{2t} = 1/\sqrt{t} + \mathcal{O}(t^{-3/2})$, and 
\[
\widetilde{W}(y_*) = - \frac{2}{3} \sqrt{t} - \frac{\gamma}{2} - \frac{\gamma^2}{4} t^{-1/2} - \frac{\gamma^3}{12 t} + \mathcal{O}(t^{-{3/2}})\ . 
\label{eq:expandwtilde1}
\]
Performing the Gaussian integral over $y$, the perturbative part of partition function becomes
\[
Z_{\text{pert}}(t) 
= 
\sqrt{ \frac{ 2\pi }{ \sqrt{\gamma^2 + 4t} } }
\sqrt{ \frac{\pi}{-  y_*} } \ e^{-t \widetilde{W}(y_*)} + \cdots\ .
\label{eq:Zpert}
\]
Therefore, the perturbative part of free energy becomes
\[
F_{\text{pert}} (t) 
&= - \log Z_{\text{pert}}(t) 
=  t\widetilde{W}(y_*) + \log \sqrt{t} +  \mathcal{O}(t^0)\notag \\
&= - \frac{2}{3} t^{3/2} - \frac{\gamma}{2} t - \frac{\gamma^2}{4} t^{1/2} + \frac{1}{2} \log t + \mathcal{O}(t^0)\ . 
\label{eq:freeenergy}
\]
The string susceptibility exponent $\gamma_{str}$ is equivalent to that of pure BPs for $t \to \infty$, i.e. $\gamma_{str} = \frac{1}{2}$.

Alternatively, we solve the differential equation (\ref{eq:u}) by perturbations for large $t$, which yields
\[
u_{\text{pert}} (t) = 
\pm t^{1/2} + \frac{\gamma}{2} \pm \frac{\gamma^2}{8} t^{-1/2} - \frac{1}{2} t^{-1} \pm \left( \frac{16 \gamma - \gamma^4}{128} \right) t^{-3/2} + \mathcal{O}(t^{-5/2})\ , 
\label{eq:upert1}
\]
where $u_{\text{pert}}$ means the perturbative part of $u$. 
The signature cannot be fixed without an external input. From Eq.~(\ref{eq:freeenergy}), we fix the signature which yields
\[
u_{\text{pert}} (t) = 
t^{1/2} + \frac{\gamma}{2} + \frac{\gamma^2}{8} t^{-1/2} - \frac{1}{2} t^{-1} + \left( \frac{16 \gamma - \gamma^4}{128} \right) t^{-3/2} + \mathcal{O}(t^{-5/2})\ .
\label{eq:upert2}
\]
As a result the perturbative part of free energy has the following expansion for large $t$
\[
F_{\text{pert}}(t) = - \frac{2}{3} t^{3/2} - \frac{\gamma}{2} t - \frac{\gamma^2}{4} t^{1/2} + \frac{1}{2} \log t + \mathcal{O}(t^0)\ . 
\label{eq:fpert}
\]
This coincides with the one obtained by the saddle-point approximation (\ref{eq:freeenergy}).

\subsection{Non-perturbative effects} 
\label{sec:nonperturbativeeffects}

Following Ref.~\cite{Hanada:2004im}, we decompose $u(t)$ into the perturbative part $u_{\text{pert}}$ and the non-perturbative part $\Delta$
\[
u(t) = u_{\text{pert}}(t) + \Delta (t)\ . 
\label{eq:udecomposition}
\]
Applying the decomposition above, the differential equation (\ref{eq:u}) can be recast as
\[
&\Delta''(t) 
+ \left( 3 (u_{\text{pert}}(t) + \Delta (t)) - \gamma \right) \Delta'(t) 
+ \left( 
3 u'_{\text{pert}}(t) + 3u^2_{\text{pert}}(t) - 2\gamma u_{\text{pert}}(t) -t
\right) \Delta (t)\notag \\
&\ \ \ 
+ \left( 3 u_{\text{pert}}(t) - \gamma\right) \Delta^2(t)
+\Delta^3(t) = 0\ , 
\label{eq:D}
\]
where we have used the fact that $u_{\text{pert}}$ satisfies the differential equation (\ref{eq:u})
\[
u''_{\text{pert}}(t) + (3u_{\text{pert}}(t)-\gamma)u'_{\text{pert}}(t) +u^3_{\text{pert}}(t) - \gamma u^2_{\text{pert}}(t) -tu_{\text{pert}}(t) - \frac{1}{2} = 0\ . 
\label{eq:diffupert}
\]
When $t\gg 1$, Eq.~(\ref{eq:D}) becomes
\[
\Delta''(t) + 3\sqrt{t} \Delta'(t) + 2t \Delta (t) + 3\sqrt{t} \Delta^2(t) + \Delta^3(t) = 0\ . 
\label{eq:D2}
\]
Assuming that $\Delta (t)$ is exponentially small when $t\gg 1$, i.e. 
\[
\Delta (t) = C e^{-h(t)}\ , 
\label{eq:Dassump}
\]
where $C$ is a constant, 
Eq.~(\ref{eq:D2}) can be approximated as 
\[
\Delta''(t) + 3\sqrt{t} \Delta'(t) + 2t \Delta (t) = 0\ . 
\label{eq:D3}
\]
Plugging Eq.~(\ref{eq:Dassump}) into Eq.~(\ref{eq:D3}), we obtain
\[
h''(t) - (h'(t))^2 + 3\sqrt{t} h'(t) -2t=0\ . 
\label{eq:hdiff}
\]
The leading solution is
\[
h(t) = \frac{2}{3} t^{3/2} -  \frac{1}{2}  \log t + \frac{1}{6} t^{-3/2} + \mathcal{O} (t^{-3})\ ,
\label{eq:hsol}
\]
which yields
\[
\Delta (t) &= C e^{-h(t)} = C \sqrt{t} \exp \left( - \frac{2}{3} t^{3/2} + \mathcal{O}(t^{-3/2}) \right)\ .
\label{de:Dsol}
\]
As a result, the free energy becomes
\[
F(t) 
= F_{\text{pert}} (t) + C e^{-S_{\text{inst}}(t)}  + \cdots\ ,
\ \ \ \text{where}\ \ \ 
S_{\text{inst}}(t) = \frac{2}{3} t^{3/2}\ . 
\label{eq:Fdecomposition}
\]

\subsection{Loop functions}
\label{sec:loopfunctions}
 
The continuum partition functions, (\ref{eq:IN1}) and (\ref{eq:zt}), are related as
\[
I_{N=1} (\lambda, g_s) = g^{2/3}_s Z(t)\ , \ \ \ \text{where}\ \ \ 
Z(t) 
= \int_D dxdy\ 
e^{y x^2 - t y + \frac{1}{3} y^3 + \frac{\gamma}{2} y^2 }\ , \ \ \ t= \frac{\lambda}{g^{2/3}_s}\ . 
\label{eq:relZI}
\]
Following the same procedure proposed in Ref.~\cite{Ambjorn:2009fm}, 
we calculate a disk amplitude that includes contributions from all genera
\[
\widetilde{w}(z) 
&= 
\frac{1}{I_{N=1} (\lambda, g_s)}
\int d\widetilde{x}d\widetilde{y}\  
\frac{1}{z - \widetilde{y}}\
e^{\frac{1}{g_s} \left( \widetilde{x}^2 \widetilde{y} - \lambda \widetilde{y} + \frac{1}{3} \widetilde{y}^3 + \frac{\gamma}{2} g^{1/3}_s \widetilde{y}^2 \right)}\notag \\
&= 
\frac{1}{\sqrt{\lambda} Z(t)} 
\int dxdy\ 
\frac{1}{\zeta - t^{-1/2} y}\ 
e^{y x^2 - t y + \frac{1}{3} y^3 + \frac{\gamma}{2} y^2 } \ \ \ (z = \sqrt{\lambda} \zeta)\notag \\
&= 
\frac{1}{\sqrt{\lambda} Z(t)} 
\int dxdy\ \int^{\infty}_0 d\alpha\ e^{-( \zeta - t^{-1/2} y )\alpha}\ 
e^{y x^2 - t y + \frac{1}{3} y^3 + \frac{\gamma}{2} y^2 }\notag \\
&= 
\frac{1}{\sqrt{\lambda}} \int^{\infty}_0 d\alpha\ e^{-\zeta \alpha}\ 
\frac{Z(t - t^{-1/2} \alpha)}{Z(t)}\notag \\
&=
\int^{\infty}_{0}d\ell\ e^{- z\ell}\ 
\frac{Z(t - t^{-1/2} \sqrt{\lambda} \ell)}{Z(t)}\ \ \ (\alpha = \sqrt{\lambda} \ell)\ ,
\label{eq:nonpertwz}
\]
where we have used Eq.~(\ref{eq:relZI}). 
From the calculation above, we can read off the following macroscopic loop function
\[
w(\ell) 
= \frac{Z (t - t^{-1/2} \sqrt{\lambda} \ell)}{Z(t)}\ . 
\label{eq:nonpertwl}
\]
This loop function includes contributions from all genera in a non-perturbative manner.

Generalizing the discussion above, one obtains
\[
\widetilde{w}(z_1, z_2, \cdots, z_n) 
&= 
\frac{1}{I_{N=1} (\lambda, g_s)}
\int d\widetilde{x}d\widetilde{y}\  
\prod^{n}_{i=1} \frac{1}{z_i - \widetilde{y}}\
e^{\frac{1}{g_s} \left( \widetilde{x}^2 \widetilde{y} - \lambda \widetilde{y} + \frac{1}{3} \widetilde{y}^3 + \frac{\gamma}{2} g^{1/3}_s \widetilde{y}^2 \right)}\notag \\
&= 
\int^{\infty}_{0}d\ell_1 \ e^{- z_1 \ell_1} \cdots \int^{\infty}_{0}d\ell_n \ e^{- z_n \ell_n}\
\frac{Z (t - t^{-1/2} \sqrt{\lambda} (\ell_1 + \ell_2 + \cdots + \ell_n))}{Z(t)}\ ,
\label{eq:nonpertmultiwz}
\]
which yields
\[
w(\ell_1, \ell_2, \cdots , \ell_n) 
= w(\ell_1 + \ell_2 + \cdots + \ell_n)
= \frac{Z (t - t^{-1/2} \sqrt{\lambda} (\ell_1 + \ell_2 + \cdots + \ell_n))}{Z(t)} \ . 
\label{eq:nonpertmultiwl}
\]

Let us find the equation that $w(\ell)$ satisfies and that reduces to the Wheeler{\textendash}DeWitt equation of $2$D causal dynamical triangulations (CDT) at $g_s=0$. 
From Eq.~(\ref{eq:diffe}), one obtains
\[
&Z''(t - t^{-1/2} \sqrt{\lambda} \ell) 
-\gamma Z' (t - t^{-1/2} \sqrt{\lambda} \ell)
- (t - t^{-1/2} \sqrt{\lambda} \ell) Z (t - t^{-1/2} \sqrt{\lambda} \ell) \notag \\
&\ \ \ 
- \frac{1}{2} \iint_{D}  dxdy \ 
\frac{1}{y}\
e^{yx^2  - (t - t^{-1/2} \sqrt{\lambda} \ell) y  + \frac{1}{3} y^3 + \frac{\gamma}{2} y^2 }= 0\ ,
\label{eq:diffZ2}
\]
which can be recast as 
\[
&\left(
- \frac{d^2}{d\ell^2} 
+ \lambda 
- g_s \ell 
- \gamma g^{1/3}_s \frac{d}{d\ell} 
\right) 
w(\ell)
= -\frac{g^{2/3}_s}{2 Z (t)} 
\iint_{D}  dxdy \ 
\frac{1}{y}\
e^{yx^2  - (t - t^{-1/2} \sqrt{\lambda} \ell) y  + \frac{1}{3} y^3 + \frac{\gamma}{2} y^2}
\ . 
\label{eq:diffZ3}
\]
Differentiating with respect to $\ell$ yields
\[
\frac{d}{d\ell}
\left[
\left(
- \frac{d^2}{d\ell^2} 
+ \lambda 
- g_s \ell 
- \gamma g^{1/3}_s \frac{d}{d\ell} 
\right) w(\ell) 
\right]
&= - \frac{g_s}{2} w(\ell) \notag \\
\Leftrightarrow\ \ \ 
\left(
- \frac{d^2}{d\ell^2} 
+ \lambda 
- g_s \ell 
- \gamma g^{1/3}_s \frac{d}{d\ell} 
\right) w(\ell) 
&= - \frac{g_s}{2} \int d\ell\ w(\ell)\ . 
\label{eq:intdiff0}
\]
Multiplying by $\ell$, one obtains
\[
\left(
- \ell \frac{d^2}{d\ell^2} 
+ \lambda \ell 
- g_s \ell^2 
- \gamma g^{1/3}_s \ell \frac{d}{d\ell} 
\right) w(\ell) 
+ \frac{g_s \ell}{2} \int d\ell\ w(\ell)
= 0 
\ . 
\label{eq:intdiff}
\]
This integro-differential equation recovers the Wheeler{\textendash}DeWitt equation of $2$D CDT at $g_s = 0$
\[
\left(
- \ell \frac{d^2}{d\ell^2} 
+ \lambda \ell 
\right) w(\ell) 
= 0\ . 
\]
One can rewrite the integro-differential equation as the following third-order linear differential equation
\[
w'''(\ell) + \gamma g^{1/3}_s w''(\ell) - (\lambda - g_s \ell) w'(\ell) + \frac{g_s}{2} w(\ell) = 0\ . 
\label{eq:thirddiffwl}
\]

The Wheeler{\textendash}DeWitt equation of generalized CDT that includes contributions from all genera is known \cite{Ambjorn:2009fm, Ambjorn:2009wi}
\[
\left(
- \ell \frac{d^2}{d\ell^2} 
+ \lambda \ell 
- g_s \ell^2 
\right) w(\ell) 
= 0\ . 
\label{eq:wdwgcdt}
\]
In comparison with the equation above, the Wheeler{\textendash}DeWitt equation we have obtained (\ref{eq:intdiff}) 
additionally includes the term proportional to $\gamma$ associated with the critical Ising spins, 
and the integral over $\ell$ that originates with the Gaussian integral over $x$.

\section{Stochastic process}
\label{sec:stochastic}

The time appeared in the string filed theories in the temporal gauge for non-critical strings corresponds to the fictitious time in the stochastic quantization \cite{Ikehara:1994xs}\footnote{
The relation to the stochastic quantization was also pointed out from a different point of view in Ref.~\cite{Jevicki:1993rr}. 
}.  
As in the case of string field theory for generalized CDT, its proper time can be identified with the fictitious time in the stochastic quantization \cite{Ambjorn:2009wi}. 

Since in Section \ref{sec:sft}, we have constructed a string field theory for the continuum limit of branched polymers (BPs) with loops coupled to the critical Ising model, 
we wish to derive the quantum Hamiltonian by identifying the time in our string field theory with the fictitious time in the stochastic quantization. 

Let us consider the stochastic variable satisfying the Langevin equation
\[
\frac{d \widetilde{y} (\tau)}{d \tau}
= - f \left(\widetilde{y} (\tau) \right) + \nu (\tau)\ , 
\label{eq:Langevin}
\]
where $\tau$ is a fictitious time, and $\nu (\tau)$ a Gaussian noise with a probability distribution functional of the form
\[
\rho (\nu) = \frac{e^{- \frac{1}{2\Omega} \int d \tau\ \nu^2(\tau)}}{ \int \prod_{\tau} d\nu(\tau)\ e^{- \frac{1}{2\Omega} \int d \tau\ \nu^2(\tau)} }\ . 
\label{eq:rho}
\]
Writing the stochastic variable subject to the initial condition, $\widetilde{y} (0) = \widetilde{y}_0$, as $\widetilde{y}(\tau ; \widetilde{y}_0)$, 
one can introduce the probability distribution function for a ``particle'' moving from $\widetilde{y}_0$ to $\widetilde{y}$
\[
P (\widetilde{y}, \widetilde{y}_0 ; \tau)
= \left\langle \delta \left( \widetilde{y} - \widetilde{y}(\tau ; \widetilde{y}_0) \right) \right\rangle_{\nu}\ , 
\label{eq:P}
\]
where the expectation value $\langle \cdot \rangle_{\nu}$ is evaluated by the distribution (\ref{eq:rho}). 
It is known that the probability distribution function (\ref{eq:P}) satisfies the Fokker-Planck equation
\[
\frac{\partial P (\widetilde{y}, \widetilde{y}_0 ; \tau)}{\partial \tau} 
= \frac{\partial}{\partial \widetilde{y}}
\left(
\frac{1}{2} \Omega \frac{ \partial P (\widetilde{y}, \widetilde{y}_0 ; \tau) }{ \partial \widetilde{y} } 
+ f(\widetilde{y}) P (\widetilde{y}, \widetilde{y}_0 ; \tau)
\right)\ . 
\label{eq:FP}
\]
From the Fokker-Planck equation, one can write $P (\widetilde{y}, \widetilde{y}_0 ; \tau )$ as the matrix element
\[
P (\widetilde{y}, \widetilde{y}_0 ; \tau )
=  \left\langle 
e^{- \tau \widetilde{H}^{\dagger}} \widetilde{y}\ \biggl{|}\ \widetilde{y}_0 
\right\rangle
= \left\langle 
 \widetilde{y}\ \biggl{|}\ 
 e^{- \tau \widetilde{H}}
 \biggl{|}\ 
 \widetilde{y}_0 
\right\rangle\ , 
\label{eq:Pmm}
\]
where $\widetilde{H}$ is the ``Hamiltonian'' operator given by
\[
\widetilde{H} = - \frac{\Omega}{2} \frac{ \partial^2 }{ \partial \widetilde{y}^2_0 } 
- \frac{\partial}{\partial \widetilde{y}_0}\ f(\widetilde{y}_0)\ .  
\label{eq:Htilde}
\]
If we introduce 
\[
\widetilde{G} (\widetilde{y}_0, \widetilde{y}; \tau) := \frac{\partial}{\partial \widetilde{y}_0} P ( \widetilde{y}, \widetilde{y}_0 ; \tau )\ ,  
\]
the function $\widetilde{G} (\widetilde{y}_0, \widetilde{y}; \tau)$ satisfies 
\[
\frac{\partial \widetilde{G} (\widetilde{y}_0, \widetilde{y} ; \tau)}{\partial \tau} 
= \frac{\partial}{\partial \widetilde{y}_0}
\left(
\frac{1}{2} \Omega \frac{ \partial \widetilde{G} (\widetilde{y}_0, \widetilde{y} ; \tau) }{ \partial \widetilde{y}_0 } 
- f(\widetilde{y}_0) \widetilde{G} (\widetilde{y}_0, \widetilde{y} ; \tau)
\right)\ . 
\label{eq:FP2}
\]

Let us rewrite the continuum partition function (\ref{eq:IN1}) as 
\[
I_{N=1} (\lambda, g_s) 
= \int d\widetilde{x}d\widetilde{y}\  e^{\frac{1}{g_s} \left( \widetilde{x}^2 \widetilde{y} - \lambda \widetilde{y} + \frac{1}{3} \widetilde{y}^3 + \frac{\gamma}{2} g^{1/3}_s \widetilde{y}^2 \right)} 
=  \sqrt{ - \pi g_s }
\int d\widetilde{y}\ 
e^{\frac{1}{g_s} S(\widetilde{y})}\ , 
\label{eq:IS}
\]
where
\[
S(\widetilde{y})
=  -\lambda \widetilde{y} + \frac{1}{3} \widetilde{y}^3 + \frac{\gamma}{2} g^{1/3}_s \widetilde{y}^2 - \frac{g_s}{2} \log |\widetilde{y}|\ . 
\label{eq:defS}
\]
Following the idea proposed by Ref.~\cite{Ambjorn:2009wi}, if we choose
\[
f (\widetilde{y}_0) = \frac{\partial S(\widetilde{y}_0)}{\partial \widetilde{y}_0}\ , 
\label{eq:defdS}
\]
Eq.~(\ref{eq:FP2}) yields
\[
\frac{\partial \widetilde{G} (\widetilde{y}_0, \widetilde{y} ; \tau)}{\partial \tau} 
= \frac{\partial}{\partial \widetilde{y}_0}
\left(
\frac{1}{2} \Omega \frac{ \partial \widetilde{G} (\widetilde{y}_0, \widetilde{y} ; \tau) }{ \partial \widetilde{y}_0 } 
- \left( 
-\lambda + \widetilde{y}^2_0 + \gamma g^{1/3}_s \widetilde{y}_0 - \frac{g_s}{2}\frac{1}{\widetilde{y}_0}
\right) \widetilde{G} (\widetilde{y}_0, \widetilde{y} ; \tau)
\right)\ . 
\label{eq:FP3}
\]
Let us introduce the Laplace transform
\[
\widetilde{G} (\widetilde{y}_0, \widetilde{y} ; \tau) 
= \int^{\infty}_0 d\ell_0 \int^{\infty}_0 d\ell\ e^{-\widetilde{y}_0 \ell_0} e^{-\widetilde{y} \ell}\
G(\ell_0, \ell ; \tau)\ .
\label{eq:LapG}
\]
Performing the inverse Laplace transform of Eq.~(\ref{eq:FP3}), for $\ell_0 > 0$, one obtains
\[
\frac{\partial G(\ell_0, \ell ; \tau)}{\partial \tau} 
= - HG (\ell_0, \ell ; \tau)\ , 
\label{eq:FP4}
\]
where 
\[
HG (\ell_0, \ell ; \tau) 
= \left( 
- \ell_0 \frac{\partial^2}{\partial \ell^2_0} + \lambda \ell_0 - \frac{\Omega}{2} \ell^2_0 - \gamma g^{1/3}_s \ell_0 \frac{\partial}{\partial \ell_0}
\right) G (\ell_0, \ell ; \tau)
+ \frac{g_s}{2} \ell_0 \int^{\ell_0}_0 d\ell'\ G(\ell', \ell ; \tau)\ . 
\]
Introducing a loop function
\[
w(\ell_0) 
= \int^{\infty}_{0}d\tau\ 
G(\ell_0 , \ell=0 ; \tau)\ ,  
\label{eq:wl0}
\]
and the boundary conditions
\[
G(\ell_0, \ell ; \tau=0) = \delta (\ell - \ell_0)\ , \ \ \ 
G(\ell_0, \ell ; \tau=\infty) = 0\ , 
\label{eq:boundarycond}
\]
one can integrate Eq.~(\ref{eq:FP4}), which yields  
\[
H w(\ell_0) = 
\left( 
-\ell_0 \frac{\partial^2}{\partial \ell^2_0} 
+ \lambda \ell_0 - \frac{\Omega}{2} \ell^2_0 
- \gamma g^{1/3}_s \ell_0 \frac{\partial}{\partial \ell_0}
\right) w (\ell_0)
+ \frac{g_s}{2} \ell_0 \int^{\ell_0}_0 d\ell'\ w(\ell')
 = 0\ . 
 \label{eq:WdW2}
\]
When we set $\Omega = 2 g_s$, Eq.~(\ref{eq:WdW2}) coincides with the Wheeler{\textendash}DeWitt equation (\ref{eq:intdiff}). 
In our previous argument, we cannot fix the overall sign of the Hamiltonian, but the derivation based on the stochastic process somehow fixes the overall sign of the Hamiltonian.

\section{Summary and discussions}
\label{eq:discussions}

We have investigated the continuum limit of branched polymers (BPs) with loops coupled to the critical Ising model at the zero temperature from various point of view. 

In terms of the continuum two-matrix model, we have derived the loop equation, and proposed a string field theory that reproduces the loop equation as the Dyson{\textendash}Schwinger equation. 
In the continuum two-matrix model, the term proportional to $\gamma$ characterizes the divergent fluctuations of spin variables.  

Setting $N=1$ in the continuum two-matrix model, the matrix integral turns to a two-dimensional integral, and we have found an integration domain that makes the two-dimensional integral converge, 
which defines a non-perturbative partition function. 
We have shown that the non-perturbative partition function satisfies a third-order linear differential equation, whereas the partition function in the continuum limit of pure BPs with loops obeys the Airy equation. 
If we formally set $\gamma = 0$ in the third-order linear differential equation, the solutions are given by products of two Airy functions, which implies that the constant $\gamma$ represents the effects of the critical Ising model.   

We have calculated the free energy, and in the large-$t$ limit where the loops are suppressed, we have read off the string susceptibility exponent $\gamma_{str}$, 
which is the same as that of BPs, i.e. $\gamma_{str} = 1/2$. In the free energy, the constant $\gamma$ appears in higher orders in the loop expansion, 
and therefore, the existence of loops would be important to see some non-trivial effects originated with the divergent spin fluctuations.  

We have also derived the Wheeler{\textendash}DeWitt equation, an integro-differential equation, whose solution is a non-perturbative loop function including contributions from all genera.  

By identifying the time in the string field theory as the fictitious time in the stochastic quantization, 
we have derived a quantum Hamiltonian and the corresponding Wheeler{\textendash}DeWitt equation that is precisely equivalent to the Wheeler{\textendash}DeWitt equation obtained from the  third-order linear differential equation. 
The quantum Hamiltonian obtained is not bounded from below, which is the same as that of generalized CDT, or equivalently, the continuum theory for pure BPs with loops. 

In summary, we have constructed firm tools to understand the continuum limit of BPs with loops coupled to the critical Ising model, which is very nice. 
 However, some interesting physics originating from quantum criticality, as well as its understanding based on quantum gravity, are not yet fully clear at present. 
 These would provide intriguing directions for future study.

\section*{Acknowledgement}
We would like to thank  
Timothy Budd, 
Masafumi Fukuma, 
Tsunehide Kuroki, 
Takuya Matsumoto, 
Shinsuke Nishigaki, 
Jun Nishimura, 
Hidehiko Shimada, 
Hiroki Takada, 
and
Osami Yasukura 
for valuable discussions. 
YS gratefully acknowledges the kind hospitality of Radboud University, Nijmegen, where part of this work was carried out. 
The work of YI was supported by JST, the establishment of university fellowships towards the creation of science technology innovation, Grant Number JPMJFS2129. 
The work of YS was partially supported by JSPS KAKENHI Grant Number 25K07318.



\end{document}